\documentclass[conference]{IEEEtran}

\usepackage{color}
\usepackage{soul}
\setulcolor{red}
\sethlcolor{yellow}

\def\@viiipt{8}
\def\squareforqed{\hbox{\rlap{$\sqcap$}$\sqcup$}}
\def\qed{\ifmmode\squareforqed\else{\unskip\nobreak\hfil
\penalty50\hskip1em\null\nobreak\hfil\squareforqed
\parfillskip=0pt\finalhyphendemerits=0\endgraf}\fi}

\usepackage{graphicx, amssymb, subfigure, setspace}
\usepackage{times}
\usepackage{algorithmicx}
\usepackage{algpseudocode}
\usepackage[ruled]{algorithm}
\usepackage{amsmath}
\usepackage{wrapfig}
\usepackage{textcomp}

\usepackage[long]{optional}

\usepackage{comment}
\usepackage{bm}
\newtheorem{lemma}{Lemma}
\newtheorem{theorem}{Theorem}

\usepackage{arydshln}

\newcommand\eat[1]{}
\newcommand\eatForMini[1]{}

\begin{document}

\title{An Auction Approach to Spectrum Management in HetNets}
\author{\IEEEauthorblockN{
Linquan Zhang\IEEEauthorrefmark{1}, Zongpeng Li\IEEEauthorrefmark{1},
Chuan Wu\IEEEauthorrefmark{2}}
\IEEEauthorblockA{\IEEEauthorrefmark{1}
Department of Computer Science, University of Calgary, 
\{linqzhan,zongpeng\}@ucalgary.ca}
\IEEEauthorblockA{\IEEEauthorrefmark{2}
 Department of Computer Science, The University of Hong
Kong, cwu@cs.hku.hk}
}

\maketitle

\begin{abstract}
The growing demand in mobile Internet access calls for high capacity and energy efficient cellular access with better cell coverage. The in-band relaying solution, proposed in LTE-Advanced, improves coverage without requiring additional spectrum for backhauling, making its deployment more economical and practical. However, in-band relay without careful management incurs low spectrum utilization and reduces the system capacity. We propose auction-based solutions that aim at dynamic spectrum resource sharing, maximizing the utilization of precious spectrum resources. We first present a truthful auction that ensures a theoretical performance guarantee in terms of social welfare. Then in an extended system model that focuses on addressing the heterogeneity of resource blocks, we design a more practical auction mechanism. We implement our proposed auctions under large scale real-world settings. Simulation results verify the efficacy of proposed auctions, showing improvements in both cell coverage and spectrum efficiency.    
\end {abstract}

\section{introduction}
\label{sec:intro}

Bandwidth-intensive mobile applications, such as video streaming, cloud-based services and social media apps, have gained momentum in cellular networks. Consequently there is an escalating demand for cellular data service with higher efficiency, larger capacity and better cell coverage. Cellular operators keep upgrading their networks and deploying new equipments to cater for such a growing demand. For example, Rogers, a major cellular service provider in Canada, recently launched their LTE-Advanced (LTE-A) service in 12 major Canadian markets \cite{lte-a-wiki}. 

As high as the traction to expand the LTE market, there is a trend to deploy more base stations for denser coverage in a geographical area. It is intended to improve signal quality in general, but comes with the cost of increased radio interference between neighbouring cells. The problem exacerbates when many users locate at the edge, rather than around the center, of a macrocell site \cite{SmallCellHetNet}. Furthermore, high frequencies employed by LTE leads to quick signal strength degradation during propagation. The situation further deteriorates with the introduction of mmWave spectrum, which is planned for the upcoming 5G communication \cite{intro5g}. Radio signals are more easily attenuated by hills, buildings, foliage and even heavy rain, making cellular communication less reliable in certain outdoor and indoor scenarios \cite{indoor-wireless, mmWave-challenges}.   

The problem could be alleviated by a heterogeneous network infrastructure (HetNet). The HetNet is a wireless network exploiting multiple types of access nodes, including macrocells, picocells and femtocells, to to improve coverage in a given area and to offer users better experience. A macrocell has the widest coverage, and serves as the main pole connecting to distant areas. Femtocells cover a much smaller range, and are typically deployed to enhance signals in compact space. Different access nodes can be deployed in a flexible combination to fit the needs of varying application scenarios such as office buildings, homes, shopping malls and subway stations \cite{SmallCellHetNet, hetnet-wiki}. 

The HetNet solution is adopted in the LTE-A standard, where Relay Nodes (RNs) are introduced to provide small cell coverage at cell edges \cite{lte-a-wiki, lte-a-3gpp}. Different from current picocells and femtocells that use fibre-based backhauls, RNs in LTE-A are connected to the macrocell via a radio interface. It is possible to deploy RNs everywhere without installing fibre or broadband infrastructure in advance. RNs hence have the potential of flexibly and substantially improve LTE-A cell coverage. Due to the use of in-band wireless connection as the backhaul, RNs may cause interference with the macrocell. The situation may become worse when a large number of RNs and user equipments (UEs) are served directly by the same donor {\em eNodeB} within the macrocell.  

A possible solution to the interference problem caused by RNs is time-sharing using a pre-determined, static inter-cell interference coordination scheme \cite{femtocells}. However, such static sharing is not flexible and may lead to resource under-utilization under time-varying workload in the RNs' small cells.
Furthermore, RNs can be installed by the users themselves, instead of mobile network operators, to improve coverage at user locations such as offices, homes, farms, and stores in large shopping malls. A coordination based solution may be vulnerable because users have no incentive to provide truthful information to the centralized controller, when falsified information may help selfish users to obtain extra benefits.

Compared with static spectrum sharing, auction based dynamic sharing can promptly adapt to network dynamics. 
The auction approach is further efficient in allocating resources to where they are most valued. Truthfulness is a key property pursued in auction mechanism design, precluding deceitful bids from strategical bidders driven by their own interests. The celebrated Vickrey-Clarke-Groves (VCG) auction is known to guarantee both truthfulness and economic efficiency in terms of social welfare maximization. However, VCG auctions require solving the underlying winner determination problem (WDP) multiple times. The underlying WDP is proven NP-hard in many real world auction scenarios, including the one in this work. Hence the VCG auction is impractical to apply directly, given its computational complexity. In the performance evaluation section, we show that solving the WDP optimally in a HetNet is impossible within in tolerable time, even using a state-of-the-art integer program solver {\tt cplex} \cite{cplex}. 

As a result, we resort to efficient approximation algorithms that run in polynomial time to conduct spectrum allocation in an LTE-A HetNet. Along with a judiciously designed payment rule, the approximation allocation algorithm can be converted to a truthful auction. We focus on two HetNet models in this work, including a relaying base station model and an extended HetNet model. For the former, resource blocks are considered as homogeneous. The latter model considers heterogeneous resource blocks, in which UEs and RNs experience varying channel quality over different channels. Two approximation algorithms are designed using the primal-dual framework. Then two auctions, aiming at soliciting truthful bids from selfish users, are proposed for the two HetNet models respectively. We prove that both auctions are truthful and individual rational, and both run in polynomial time. Using a primal-dual analysis based on linear programming duality, we also prove an approximation ratio that guarantees the outcome would be nearly optimal for the relaying base station model. In particular, in practical scenarios where the total amount of allocated resource is far larger than the amount of resource requested by a single user in a single time slot, the theoretical ratio approaches $1/e$, where $e$ is the base of natural logarithm. 

Extensive simulation studies are conducted to evaluate the proposed auction mechanisms at large system scales. Simulation results exhibit close-to-optimum performance in terms of social welfare for both mechanisms, much better than suggested by the theoretical worst-case bound $1/e$. The total throughput and individual throughput are further compared, showing that the proposed mechanisms achieve relatively high system throughput. For the extended HetNet model, we also compare the proposed auction with alternative schedulers from the literature, revealing that our auction could produce higher social welfare and relatively fair resource sharing.

The rest of the paper is organized as follows. We discuss background and related work in Sec.~\ref{sec:relatedwork}. The system model is introduced in Sec.~\ref{sec:model}, with their truthful auctions designed and analyzed in Sec.~\ref{sec:simple} and Sec.~\ref{sec:extended}. Sec. \ref{sec:experiment} contains performance evaluation, and Sec.~\ref{sec:conclusion} concludes the paper.

\section{Background \&  Related Work}
\label{sec:relatedwork}

\subsection{Background of LTE and LTE-Advanced Networks}
An LTE system has two major components: the Evolved Packet Core (EPC) and the evolved NodeBs ({\em eNodeB}) or basestations \cite{lte-a-wiki, lte-a-3gpp, e-lte-a}. The EPC is composed of a range of entities, among which the Mobility Management Entity (MME) provides key control functions in the control plane, and the Home Subscriber Server (HSS) stores user information in its database. The Serving Gateway (SGW) is connected to the {\em eNodeB}s, and serves as the end point of E-UTRAN in the data plane. The Packet Gateway (PGW) provides connectivity to external packet data network. LTE-Advanced (LTE-A) \cite{lte-advanced,e-lte-a}, emerging as a major extension of LTE recently as a complete fulfilment of 4G, provides up to $3$ Gbps download link access and up to $1.5$ Gbps upload link access by incorporating i) carrier aggregation, ii) MIMO techniques, and iii) in-band relay. In particular, the relay nodes improve the cell coverage by connecting to an {\em eNodeB} via radio interface.

\noindent {\bf Resource Allocation.}
The minimum allocation unit of radio resource in LTE and LTE-A is a Resource Block (RB), whose size is $180$ kHz in frequency domain and $0.5$ ms in time domain \cite{e-lte-a}. In practice, RBs are often allocated in a group. The group size varies depending on the configuration of bandwidth. For example, for system bandwidth between $11$ and $26$ MHz, a group of two RBs form an atomic resource allocation unit.

\noindent {\bf CQI Reporting.} In LTE and LTE-A, Channel Quality Indicator (CQI) reporting plays a critical role in spectrum allocation and transmission scheduling. the CQI value ranges from $0$ to $15$ \cite{e-lte-a}. Each value indicates a supportable modulation (QPSK, 16QAM, 64QAM) and coding rate. High CQI values imply high channel quality and high data rates.

\subsection{Related work} 
A series of recent studies focus on wireless relay networks. Oyman \cite{OymanTWC10} analyzes the spectrum efficiency of opportunistic scheduling and spectrum reuse algorithms in orthogonal frequency-division multiple-access (OFDMA) cellular networks with relay stations. Our work instead studies the spectrum resource allocation using an auction approach.
Ren {\em et al.} \cite{RenTWC10} study relay transmission in the downlink of a multi-channel time-division-multiple-access (TDMA) cellular network, presenting distributed algorithms to compute power allocation for self-interested relay nodes. Huang {\em et al.} \cite{HunagJASC08} propose mechanisms to determine relay selection as well as relay power allocation in a cooperative communication network. Both works focused on the existence of Nash Equilibrium of proposed mechanisms, while our work focuses on the dominant-strategy solutions with individual rationality and good approximation ratio in social welfare.

Wireless spectrum is a scarce resource; efficient utilization of spectrum resources is key to the optimization of a wireless communication network. Spectrum auctions represent an efficient solution approach, which have been extensively studied in wireless networks. Jia {\em et al.} \cite{Jia2009} propose two spectrum auctions: an optimal VCG-type auction and a suboptimal truthful auction. However the former is computationally infeasible while the latter does not provide performance guarantee in terms of approximation ratio.  Zhu {\em et al.} \cite{Yuefei12} design a truthful auction for secondary spectrum markets. Nevertheless their design exploits a packing structure in the underlying problem while our problem is essentially a non-packing problem. 

Auctions are also widely used in cloud computing to dynamically provision computing resources to users. For example, Zhang {\em et al}. \cite{linquan-infocom14-auction} proposed a randomized auction for cloud computing using a primal-dual decomposition technique. Shi {\em et al.}~\cite{shi-sigmetrics14} further extend the randomized auction to an online scenario where user requests arrive on the fly. Yet again both solutions require the underlying WDP to be of packing type.

\section{System Model}
\label{sec:model}
We consider an LTE-A HetNet where the wireless network employs heterogeneous types of access nodes.
The system runs in a time slotted fashion. In each time slot, the HetNet system, whose controller is behind the donor {\em eNodeB} and located in the EPC, allocates available resource blocks (RBs) to the RNs as well as the UEs, while avoiding interference within the cell. Towards efficient RB allocation, the system adopts an auction-based allocation approach. Suppose that the downlink in a single donor cell consists of a donor {\em eNodeB} and a set $\mathcal{R}$ of RNs with cardinality $R$. Let $A$ be the spectrum band the {\em eNodeB} is using. The total number of RBs within a single time slot is $N$.

\begin{figure}[!htbp]
	\begin {center}
	\includegraphics[width=0.45\textwidth]{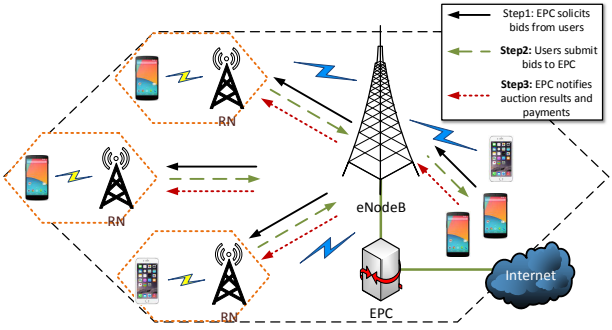}	
	\caption{Downlink resource block allocation in a HetNet.}
	\label{fig:model}
	\end {center}
\end{figure}

At each time slot, the controller in the EPC acts as the auctioneer, soliciting bids from bidders that include RNs as well as UEs, as shown in Figure \ref{fig:model}. 

\subsection{The Relaying Base Station Model}
\label{subsec:rns-model}
Our first model, the relaying BS Model, 
focuses on allocating homogeneous RBs among RNs as well as UEs based on their submitted bids, while ensuring that a sufficient number of RBs is reserved for the intra-RN communication so that all participants in the system can communicate simultaneously. Assume that RNs' areas of coverage do not overlap.
Let $\mathcal{B}=\mathcal{B}_r \cup \mathcal{B}_u$ denote the set of bidders, where $\mathcal{B}_r$ is the set of RNs and $\mathcal{B}_u$ is the set of UEs served directly by the donor {\em eNodeB}. Let $b_i = (r_{i}, v_i)$ be the bid submitted by user $i$, where $r_{i}$ is the number of RBs it requests, for which it is willing to pay up to $v_i$. $x_i$ is a binary variable indicating whether bid $b_i$ wins. We also have a $(|\mathcal{B}|+1)\times N$ matrix $A = (a_{ik})$, where $a_{ik}$ is a 0/1 variable indicating whether RB $k$ is allocated to bidder $i$ (bidder $|\mathcal{B}|+1$ denotes the resource reserved for the transmission within a relay node's coverage). At the end, each relay node $i\in\mathcal{B}_r$ receives an auction result indicating which RB it is allocated for receiving transmission from the donor {\em eNodeB} and which RB it could use to transmit data to the UEs served by the relay node.

The number of RBs allocated to a winning bidder $i$ should equal the number requested in its bid, {\em i.e.}, $\sum_{k} a_{ik} = r_{i} x_i, \forall 1 \leq i \leq |\mathcal{B}|$.

Each RB can be allocated to at most one bidder in the cell, in order to avoid interference, {\em i.e.}, $\sum_{i=1}^{|\mathcal{B}|+1} a_{i,k} \leq 1, \forall k.$

Another important constraint is that the number of RBs allocated to each relay node for downloading data from the donor {\em eNodeB} should be smaller than that reserved for the inner communication between RNs and their UEs, {\em i.e.}, the system should reserve a sufficient number of RBs for the intra-RN communication in order to achieve seamlessly transmission between UEs served by RNs and the donor {\em eNodeB}.

{\small
\begin{equation*}
\sum_{k} a_{i,k} \leq \sum_{k}a_{|\mathcal{B}|+1, k}, \forall i\in \mathcal{B}_r.
\end{equation*}}

Let $p_i$ and $\tilde{v}_i$ be the payment and true valuation of bidder $i$, respectively. The {\em utility} of $i$ is then:

{\small 
\begin{eqnarray*}
u_i  = \big\{\begin{array}{cc}
\tilde{v}_i - p_i & \mbox{if bidder $i$ wins}.\\
0&\mbox{otherwise}
\end{array}\big.
\end{eqnarray*}}

The {\em social welfare} of the HetNet is defined as the summation of all utilities, including the seller and the bidders. Under truthful bidding, social welfare equals $\sum_{i,j} v_i^j x_i^j$, since payments cancel themselves. The winner determination problem (WDP) is:

{\small
\begin{equation}
\label{eqn:wdp-no-cqi}
\mbox{maximize}~~~ \sum_{i} v_i x_i
\end{equation}
subject to:

\begin{eqnarray*}
\sum_{k} a_{ik} = r_{i} x_i, &\forall 1 \leq i \leq |\mathcal{B}|&(\ref{eqn:wdp-no-cqi}a)\\
\sum_{i=1}^{|\mathcal{B}|+1} a_{i,k} \leq 1, &\forall k&(\ref{eqn:wdp-no-cqi}b)\\
\sum_{k} a_{i,k} \leq \sum_{k} a_{|\mathcal{B}|+1, k}, &\forall i\in \mathcal{B}_r&(\ref{eqn:wdp-no-cqi}c)\\
a_{ik}, x_i \in \{0,1\},&\forall i,k&(\ref{eqn:wdp-no-cqi}d)\\
\end{eqnarray*}
}

The allocation result includes, for each user $i\in\mathcal{B}$: 
which RBs can be used to download data from the donor {\em eNodeB}, and which can be used to send data to UEs served by RNs.

\begin{theorem}
The optimization problem WDP ($\ref{eqn:wdp-no-cqi}$) is NP-hard.
\end{theorem}

\noindent {\em Proof:}
Consider a special case where $|\mathcal{B}_r| = 0$, then constraint $(\ref{eqn:wdp-no-cqi}c)$ can be removed. Allocating nothing to the $|\mathcal{B}|+1$ user, which is considered as RBs reserved for intra-RN communication, will not decrease the maximum $\sum_{i} v_i x_i$. So we do not differentiate the RBs. $(\ref{eqn:wdp-no-cqi}a)$ and $(\ref{eqn:wdp-no-cqi}b)$ are equivalent to $\sum_{i} r_i x_i = \sum_{i,k} a_{ik} \leq N$ in this case. WDP ($\ref{eqn:wdp-no-cqi}$) is then reduced to the classic 0-1 knapsack that is NP-hard.\qed

\subsection{Extended HetNet Model}
We further consider the case in which bids submitted by the users include their CQI information. We consider the higher-layer-configured case in an aperiodic reporting situation where users report CQI for a predetermined set of sub-bands. The size of a sub-band is typically two or more RBs. In practice, users are concerned with not only how many RBs they receive but the actual throughput achievable over those RBs, given that the RBs are heterogeneous to users. Let a bid $b_i = (\bm{c_i}, r_i, v_i)$, where $\bm{c_i}$ is a vector that contains the CQI values for all sub-bands, $r_i$ is the number of RBs requested, and $v_i$ is the maximum amount the bidder is willing to pay for a unit amount of data for the next time slot ({\em e.g.}, 1 MB). Let $x_i$ be the decision variable indicating whether bidder $i$ wins or not, and $A = (a_{ik})$ be the allocation matrix.

The constraints are similar to those in Sec. \ref{subsec:rns-model}. We next determine the social welfare. Assume bidder $i$ obtains a set of RBs $\mathcal{R}_i$, where $|\mathcal{R}_i|=r_i$, and its true CQI is $\bm{c}^*_i$. The total social welfare under truthfully bidding is $\sum_i v_i \sum_{k:a_{ik}\neq 0} \mbox{CR}_{c^*_{i,{p(k)}}} a_{ik}$,
where $p(k)$ is the subframe where RB $k$ is located, $c^*_{i,p(k)}$ is the CQI of RB $k$ for user $i$, and $\mbox{CR}_{c^*_{i,p(k)}}$ is the amount of information bits in RB $k$ for user $i$. The WDP of the extended HetNet model is:

{\small
\begin{equation}
\label{eqn:wdp-cqi}
\mbox{maximize:}~~~ \sum_{i} \Big\{ v_i \sum_{k:a_{ik}\neq 0} \mbox{CR}_{c^*_{i,p(k)}} a_{ik} \Big\}
\end{equation}
subject to:

\begin{eqnarray*}
\sum_{k} a_{ik} = r_{i} x_i, &\forall 1 \leq i \leq |\mathcal{B}|&(\ref{eqn:wdp-cqi}a)\\
\sum_{i=1}^{|\mathcal{B}|+1} a_{i,k} \leq 1, &\forall k&(\ref{eqn:wdp-cqi}b)\\
\sum_{k} a_{i,k} \leq \sum_{k} a_{|\mathcal{B}|+1, k}, &\forall i\in \mathcal{B}_r&(\ref{eqn:wdp-cqi}c)\\
a_{ik}, x_i \in \{0,1\},&\forall i,k&(\ref{eqn:wdp-cqi}d)\\
\end{eqnarray*}
}

\begin{theorem}
The optimization problem defined in WDP (\ref{eqn:wdp-cqi}) is NP-hard.
\end{theorem}

\noindent {\em Proof:} Consider the case where all CQI values are equal. Then the objective of the problem becomes:

{\small
\begin{equation*}
\sum_{i} \Big\{ v_i \sum_{k:a_{ik}\neq 0} \mbox{CR}_{c^*_{i,p(k)}} a_{ik} \Big\} = \sum_i \Big\{v_i r_{i} x_{i} \mbox{CR}_{c^*_i} \Big\} = \sum_{i} \tilde{v}_i' x_{i},
\end{equation*}}

\noindent where $\tilde{v}_i' = v_i r_{i} \mbox{CR}_{c^*}$. Then the problem reduces to the WDP (\ref{eqn:wdp-no-cqi}), which we proved to be NP-hard. \qed

\section{Auction Design for the Relaying BS Model}
\label{sec:simple}
As the underlying WDP (\ref{eqn:wdp-no-cqi}) is NP-complete, we consider its LP relaxation by relaxing the constraint $(\ref{eqn:wdp-no-cqi}d)$ to $a_{i,k}\geq 0, 1\geq x_i \geq 0$, instead of solving the integer programming problem directly. Then by introducing dual variables $\bm{\beta}, \bm{\lambda}, \bm{\rho}$ and $\bm{\xi}$ corresponding to $(\ref{eqn:wdp-no-cqi}a)-(\ref{eqn:wdp-no-cqi}c)$ and $x_i\leq 1, \forall i\in\mathcal{B}$, respectively, we formulate the dual of $(\ref{eqn:wdp-no-cqi})$ as follows:

{\small
\begin{equation}
\label{eqn:wdp-no-cqi-dual}
\mbox{minimize}~~~\sum_{k=1}^{|N|} \lambda_k + \sum_{i=1}^{|\mathcal{B}|} \xi_i
\end{equation}
subject to:

\begin{eqnarray*}
r_i \beta_i + \xi_i - v_i \geq 0& \forall i\in\mathcal{B}&(\ref{eqn:wdp-no-cqi-dual}a)\\
\lambda_k \geq \max\{\sum_{i\in\mathcal{B}_r} \rho_i, \max_{i\in\mathcal{B}} (\beta_i - \rho_i \bm{1}_{i\in \mathcal{B}_r})\}& \forall k&(\ref{eqn:wdp-no-cqi-dual}b)\\
\beta_i \hspace*{2mm}\mbox{unconstrained}, \lambda_k, \rho_i, \xi_i \geq 0&\forall i,k&(\ref{eqn:wdp-no-cqi-dual}c)\\
\end{eqnarray*}}

\noindent where $\bm{1}_{i\in \mathcal{B}_r} = 1 \mbox{ if } i\in \mathcal{B}_r; 0$ otherwise. Constraint $(\ref{eqn:wdp-no-cqi-dual}b)$ can be reformulated into the following three inequalities: 
{\small
\begin{eqnarray*}
\lambda_k-\beta_i \geq 0, \forall i\in \mathcal{B}_u, k\\
\lambda_k-\beta_i + \rho_i \geq 0, \forall i\in \mathcal{B}_r, k\\
\lambda_k -\sum_{i\in\mathcal{B}_r} \rho_i \geq 0, \forall k
\end{eqnarray*}}
Constraint (\ref{eqn:wdp-no-cqi-dual}b) does not differentiate among RBs, which can be considered identical. Furthermore, $\beta_i$ can also be substituted by $\lambda$ and $\rho_i$. The problem becomes:
{\small
\begin{equation}
\label{eqn:wdp-no-cqi-dual2}
\mbox{minimize}~~~|N| \lambda + \sum_{i=1}^{|\mathcal{B}|}\xi_i 
\end{equation}
subject to:

\begin{eqnarray*}
\hspace*{1.5mm}r_i \lambda + \rho_i \bm{1}_{i\in\mathcal{B}_r}+\xi_i - v_i \geq 0& \forall i\in\mathcal{B}&(\ref{eqn:wdp-no-cqi-dual2}a)\\
\lambda -\sum_{i\in\mathcal{B}_r} \rho_i\geq 0& &(\ref{eqn:wdp-no-cqi-dual2}b)\\
\lambda, \rho_i \geq 0&\forall i&(\ref{eqn:wdp-no-cqi-dual2}c)\\
\end{eqnarray*}}

Let $\delta =  |N|/\max_{i\in\mathcal{B}} r_i$. We design the following approximation algorithm shown in Algorithm \ref{alg:greed}. 
The first part of Algorithm \ref{alg:greed} initializes the primal and dual variables. Then we sort all bids from users in descending order of $v_i/r_i$. Next, a {\tt while} loop iteratively updates the primal and dual variables. 

The key idea behind Algorithm \ref{alg:greed} is to find a bidder $\mu$ such that $v_{\mu}/r_{\mu}$, which can be interpreted as the value of a unit-weight RB, is maximized among the remaining unallocated bidders. It then allocates the corresponding number of RBs according to the demand of $\mu$. 

The exit condition and the update rule for the dual variable $\lambda$ are designed carefully to ensure the feasibility of the primal variables, while dual variables (not necessarily feasible upon algorithm termination) can be converted to a feasible solution through the technique of dual fitting. The approximation ratio of the primal-dual algorithm can be analyzed by comparing the primal solution and the converted dual solution, establishing a bound on their ratio, and then applying weak LP duality. 

\begin{algorithm}[!htp]
\caption{ A Primal-Dual Allocation Algorithm}\label{alg:greed}
\begin{algorithmic}[1]
\State {// {\small Initialization}}
\State {\small $p = 0; t=0;$		$\mathcal{C} = \emptyset$;}
\State {\small  $\forall i\in\mathcal{B}: x_{i} = 0$, $\rho_i=0$, $\xi_i = 0$, $\lambda^0 = 1/|N|$;}
\State
\State {\small Sort $\{\frac{v_i}{r_i}|\forall i\in\mathcal{B}\}$;}
\While{\small $\mathcal{C}\neq \mathcal{B}$ AND $|N|\lambda^t \leq \exp(\delta-2)$} 
	\State $\mu = \arg\max_{i\in \mathcal{B}\setminus\mathcal{C}}\big\{\frac{v_i}{r_i} \big\}$;	
	\State $x_{\mu} = 1;$ $\xi_{\mu} =  v_{\mu}$;
	\State $p = p+v_{\mu}$; $\mathcal{C} = \mathcal{C}\cup\{\mu\};$
	\If{$\mu\in \mathcal{B}_r$}
	    \State $\rho_{\mu} = \frac{\lambda^{t}}{|\mathcal{B}_r|}$
	\EndIf
	\State $\lambda^{t+1} = \lambda^{t} \cdot \exp(\delta-2)^{r_{\mu}/(|N|-2\max_{i\in\mathcal{B}}r_i)}$;
	\State $t = t+1;$
\EndWhile
\end{algorithmic}
\end{algorithm}

The feasibility of Algorithm \ref{alg:greed} is ensured by the following theorem, which also shows that Algorithm \ref{alg:greed} runs in polynomial time. The proof is in \opt{long}{Appendix~\ref{appendix:theorem3}}\opt{short}{the technical report \cite{technical-report}}.

\begin{theorem}
\label{thm:alg1_runtime}
Algorithm \ref{alg:greed} computes a feasible solution to the primal problem in $O(|\mathcal{B}|\log |\mathcal{B}|)$.
\end{theorem}


The approximation ratio of Algorithm \ref{alg:greed} is revealed in the following theorem, whose proof is available in \opt{long}{Appendix \ref{appendix:theorem4}}\opt{short}{the technical report \cite{technical-report}}.

\begin{theorem}
\label{thm:alg1-appro}
Algorithm \ref{alg:greed} guarantees $\alpha$-approximation, where $\alpha = \frac{\delta-2}{\delta e - 2}$, $\delta =  |N|/\max_{i\in\mathcal{B}} r_i$.
\end{theorem}

Algorithm \ref{alg:greed} achieves at least an $\alpha = (\delta-2)/(\delta e -2)$ fraction of optimal social welfare. Yet it does not eliminate deceitful bids from strategic users. A carefully designed payment scheme can help the donor {\em eNodeB} rule out falsified bids, ensuring the truthfulness of the mechanism. Let $\bm{b} = (b_i, b_{-i})$ be the bids submitted by the bidders, where $b_i$ is the bid of bidder $i$ and $b_{-i}$ includes the bids of others. $P_i(b_i,b_{-i})$ is the winning probability when bidder $i$ submits bid $b_i$. The following theorem characterizes a sufficient and necessary condition of the truthfulness of an auction.

\begin{theorem}
\label{thm:truthful-iff}
\cite{Myerson} An auction with bids $\bm{b}$ and payments $\bm{\Pi}$ is truthful in expectation if and only if
\begin{enumerate}
\item $P_i(b_i)$ is monotonically non-decreasing in $b_i$, $\forall i\in\mathcal{B}$;
\item The expected payment satisfies:

{\small
\begin{equation*}
E[\Pi_i] = b_i P_i(b_i,b_{-i}) - \int_{0}^{b_i} P_i(b,b_{-i}) db, \forall i\in\mathcal{B}.
\end{equation*}}
\end{enumerate}
\end{theorem}

Assume all bids request the same numbers of RBs, and bidder $i$ wins with bidding price $v_{i}$. If other bids $v_{-i}$ remain the same, then bidder $i$ wins as well with bidding price higher than $v_{i}$, since the allocation rules in Algorithm \ref{alg:greed} is greedy.
If the bidding price is lower than $v_{i}$, then bidder $i$ might lose since other bids can be selected before $b_i$. That is, $P(v_i, v_{-i})$, the probability of user $i$ wins when submitting $(r_i, v_i)$, is non-decreasing in $v_i$. For a deterministic mechanism, there are only two possible values $0$ and $1$ for $P_i(v_i)$. A critical bidding price $v_i^*$ for each winning bidder $i$ is that 

{\small
\begin{equation*}
P_i(v_i) \left\{\begin{array}{ll}
0 & \mbox{ if } v_i<v_i^*\\
1 & \mbox{ otherwise}\\
\end{array}\right.
\end{equation*}}
Given $v_i^*$, the corresponding payment is 
{\small
\begin{equation*}
\Pi_i =  v_i P_i(v_i,v_{-i}) - \int_{v_i^*}^{v_i} P_i(v,v_{-i}) dv = v_i^*.
\end{equation*}} 

Following the definition, the critical bid can be computed via a binary search between $0$ and $v_i$ for any winning bidder $i$, such that bidder $i$ wins when $v_i\geq v_i^*$, and loses otherwise. The detailed payment rule is described in Algorithm \ref{alg:payment1}.

\begin{algorithm}[!htp]
\caption{ Payment Rules}\label{alg:payment1}
\begin{algorithmic}[1]
\State {// {\small Initialization}}
\State {\small $\forall i\in\mathcal{B}, \Pi_i = 0$;}
\State
\For{\small $i\in\mathcal{B}$}
\If{\small bidder $i$ wins, {\em i.e.}, $x_i==1$}
\State{\small $a=0, b=v_i$;}
\While{\small $(b-a)>\epsilon$}
    \State{\small Run Algorithm \ref{alg:greed} with $(\frac{a+b}{2}, v_{-i})$ as bids;}
    \If{\small bidder $i$ wins}
        \State{\small $b=\frac{a+b}{2}$;}
    \Else
        \State{\small $a = \frac{a+b}{2}$;}
    \EndIf
\EndWhile
\State{\small $\Pi_i = (b+a)/2$;}
\Else
\State{$\small \Pi_i=0$;}
\EndIf
\EndFor
\end{algorithmic}
\end{algorithm}

\begin{theorem}
The time complexity of the payment rules in Algorithm \ref{alg:payment1} is $O(|\mathcal{B}|^2\log(m/\epsilon)\log|\mathcal{B}|)$.  
\end{theorem}

\vspace{1mm}
\noindent {\em Proof}:
The {\tt for} loop is executed exactly $|\mathcal{B}|$ times. Within the {\tt for} loop. The {\tt while} loop is a binary search that has $O(\log v_i/\epsilon)$ steps. Line 8 costs $O(|\mathcal{B}|\log|\mathcal{B}|)$ time according to Theorem \ref{thm:alg1_runtime}. Therefore the total time complexity is:

{\small
\begin{equation*}
O(|\mathcal{B}|\log(v_i/\epsilon)\cdot |\mathcal{B}|\log|\mathcal{B}|)\leq O(|\mathcal{B}|^2\log(m/\epsilon)\log|\mathcal{B}|). ~~~\qed
\end{equation*}}

\begin{theorem}
\label{thm:alg1-apprx-ratio}
The allocation rule (Algorithm \ref{alg:greed}) and the payment rule (Algorithm \ref{alg:payment1}) together constitute an auction that is truthful, individual rational, and achieves $\alpha$-approximation in social welfare of the HetNet, where $\alpha = \frac{\delta-2}{\delta e-2}.$
\end{theorem}

\vspace{1mm}
\noindent {\em Proof:} As discussed, the winning probability of user $i$ is non-decreasing in the bidding price $v_i$. We then examine whether the payment rule satisfies the second condition in Theorem \ref{thm:truthful-iff}. Given a positive real number $\epsilon$, for a winning bidder $i$, we always find a payment $\Pi_i$ where $|\Pi_i-v_i^*|\leq \epsilon$, {\em i.e.}, the critical bidding price can be found within a given margin of error. Therefore the second condition is also satisfied, and the spectrum auction is truthful.

The utility of bidder $i$ under truthfully bidding is then:

{\small
\begin{equation*}
u_i = \tilde{v}_i - \Pi_i = v_i - \Pi_i.
\end{equation*}} 

Due to the payment rules in Algorithm \ref{alg:payment1}, $v_i \geq \Pi_i$, which implies that $u_i\geq 0$, {\em i.e.}, a bidder's utility will be non-negative when it bids truthfully. Thus the auction is individual rational. 

Theorem \ref{thm:alg1-appro} shows the allocation rule in Algorithm \ref{alg:greed} is a $\frac{\delta -2}{\delta e -2}$-approximation, so the auction is $\frac{\delta -2}{\delta e -2}$-approximation in system-wide social welfare as well.
\qed

\begin{figure}[!htp]
	\begin {center}
    \begin{minipage}[t]{0.49\linewidth}
    \centering
	\includegraphics[width=\textwidth]{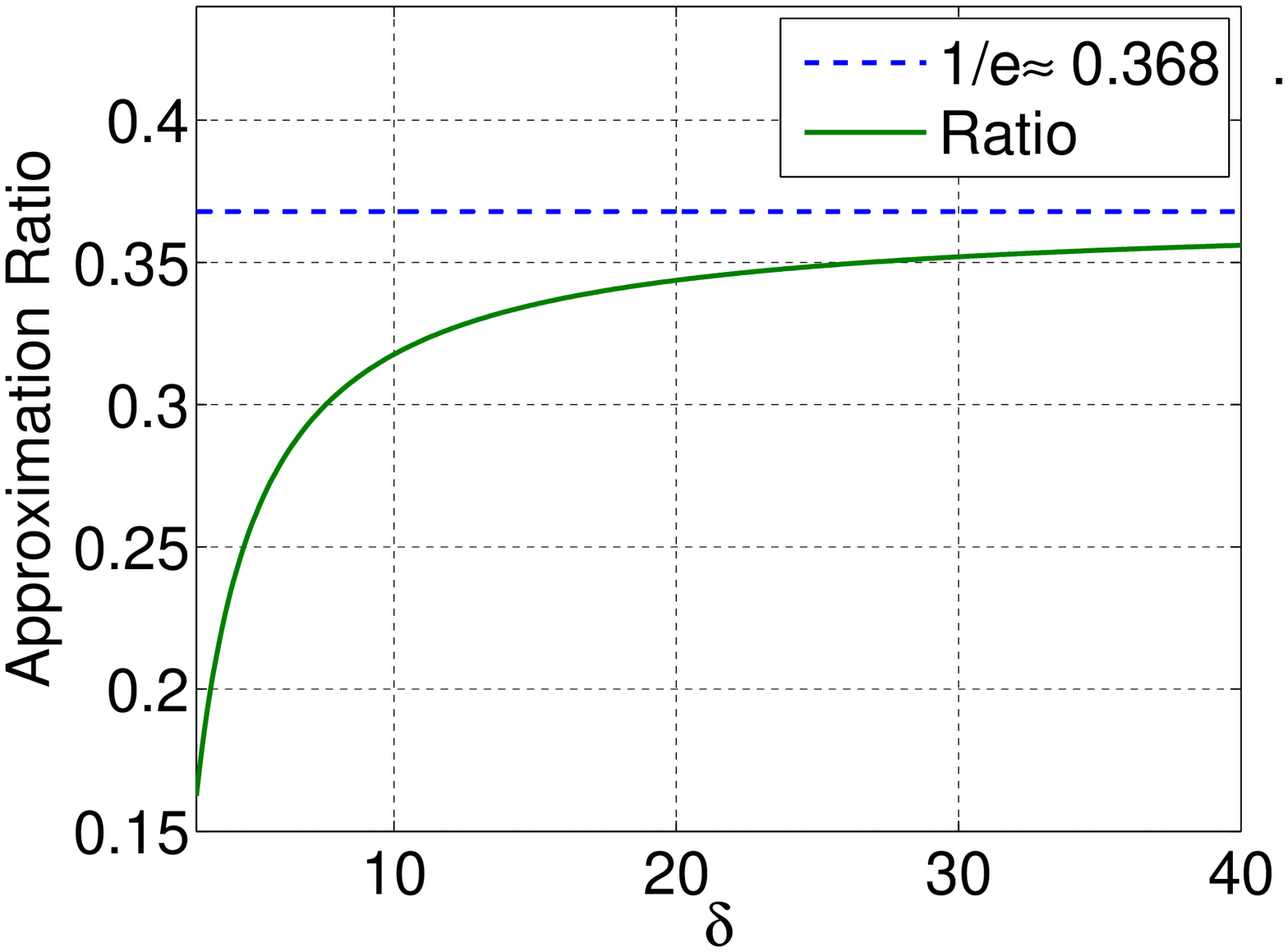}
	\caption{{\small An illustration of the approximation ratio as a function of $\delta$. The ratio approaches $1/e$ as $\delta$ increases.}}
	\label{fig:ratio}
    \end{minipage}
    \hfill
    \begin{minipage}[t]{0.49\linewidth}
    \centering
	\includegraphics[width=\textwidth]{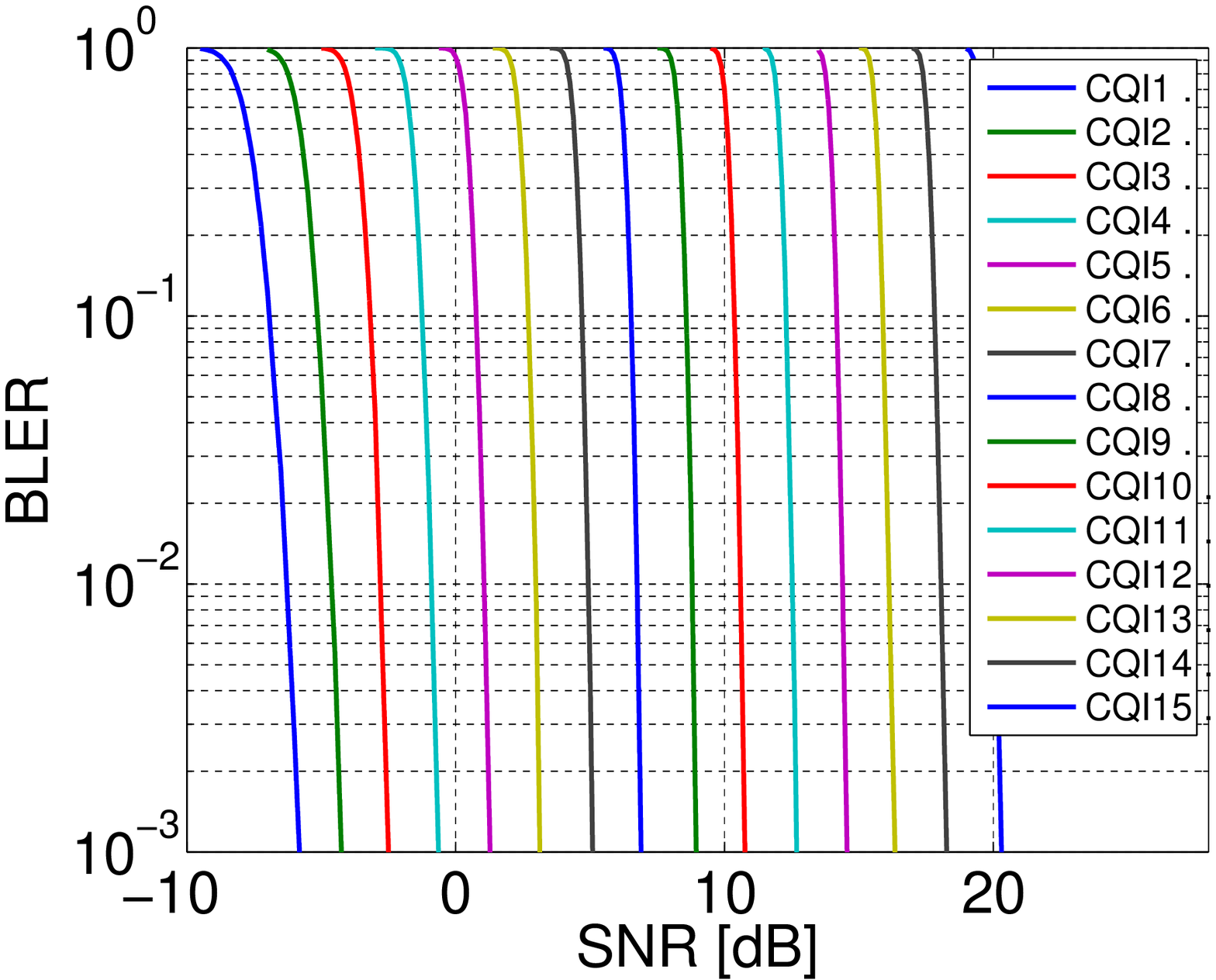}
	\caption{{\small CQI vs. Block Error Ratio. When $\mbox{CQI}_{\mbox{report}}-\mbox{CQI}_{\mbox{actual}}\geq 1$, BLER is close to 1, transmission fails.}}
	\label{fig:cqi-bler}
    \end{minipage}
	\end {center}
\end{figure}

Note that $\lim_{\delta\rightarrow\infty} \frac{\delta-2}{\delta e -2} = \frac{1}{e}$, which indicates that the ratio approaches $1/e\approx 0.368$ as $\delta$ increases. In practice, the total number of available RBs is typically much larger than the demand from a single user device, {\em i.e.}, $N\gg m=\max_{i\in\mathcal{B}}r_i$. Consequently, the ratio is approximately $0.368$. The curve of $\frac{\delta-2}{\delta e -2}$ is plotted in Figure \ref{fig:ratio}, showing that the ratio quickly approaches $0.368$ as $\delta$ increases.

\section{A Truthful Auction for Extended HetNet}
\label{sec:extended}

In LTE-A, the base station chooses coding rate as well as the modulation scheme according to the CQI reported by UEs. It may appear that strategic users could report falsified CQI values in the hope of increasing its chances of winning channels. However, reporting CQI higher than the actual value may lead to high transmission failure rates \cite{Pelechrinis14}, making effective data transmission almost impossible, as shown in Figure \ref{fig:cqi-bler}. Meanwhile, reporting lower CQI leads to lower coding rate and lower throughput even though transmission success rate becomes higher. Therefore selfish users typically have no motivation to falsify CQI reports in practice. In the following design and analysis, we assume that CQIs and number of desired RBs reported by the users are true. 

Let $R_{i,k} = v_i \mbox{CR}_{c^*_{i,p(k)}}, \forall i, k$, then the objective of the extended HetNet model is $\sum_{i,k} R_{ik} a_{ik}$. We derive the dual of (\ref{eqn:wdp-cqi}) by introducing dual variables $\tilde{\bm{\beta}}, \tilde{\bm{\lambda}}, \tilde{\bm{\rho}}$ and $\tilde{\bm{\xi}}$ corresponding to $(\ref{eqn:wdp-cqi}a)-(\ref{eqn:wdp-cqi}c)$ and $x_{i}\leq 1, \forall i\in\mathcal{B}$, respectively:

{\small
\begin{equation}
\label{eqn:wdp-cqi-dual}
\mbox{minimize}~~~\sum_{k=1}^{|N|} \tilde{\lambda}_k + \sum_{i=1}^{|\mathcal{B}|} \tilde{\xi}_{i}
\end{equation} 
subject to:
\begin{eqnarray*}
\tilde{\lambda}_k \geq \max\{\sum_{i\in\mathcal{B}_r} \tilde{\rho}_i, \max_{i\in\mathcal{B}}(\tilde{\beta}_i -\tilde{\rho}_i \bm{1}_{i\in\mathcal{B}_r} + R_{i,k})\} , &\forall k\in\mathcal{N}&(\ref{eqn:wdp-cqi-dual}a)\\ 
r_i \tilde{\beta}_i + \tilde{\xi}_i \geq 0, &\forall i\in\mathcal{B}&(\ref{eqn:wdp-cqi-dual}c)\\
\tilde{\lambda}_k\geq 0, \tilde{\rho}_i \geq 0, \tilde{\xi}_{i}\geq 0&\forall i,k&(\ref{eqn:wdp-cqi-dual}d)\\
\end{eqnarray*}}

To minimize $\sum_{k=1}^{|N|} \tilde{\lambda}_k + \sum_{i=1}^{|\mathcal{B}|} \tilde{\xi}_{i}$, we set $\tilde{\beta}_i = 0$ if $\tilde{\xi}_i = 0$;  $\tilde{\beta}_i = -\tilde{\xi}_i/r_i$ if $\tilde{\xi}_i >0$. In either case, we can write $\tilde{\beta}_i = -\tilde{\xi}_i/r_i$. Substitute $\tilde{\beta}_i$ in (\ref{eqn:wdp-cqi-dual}a) with $\tilde{\beta}_i = -\tilde{\xi}_i/r_i$, then we have 
{\small
\begin{eqnarray*}
\tilde{\lambda}_k + \tilde{\xi}_i/r_i +\tilde{\rho}_i \bm{1}_{i\in\mathcal{B}_r} \geq& R_{i,k}, & \forall i,k\\
\tilde{\lambda}_k \geq& \sum_{i\in\mathcal{B}_r} \tilde{\rho}_i, &\forall k\\
\end{eqnarray*}
}

Given the different problem structure resulting from heterogeneous RBs, we design a new greedy primal-dual algorithm in Algorithm \ref{alg:allocation2} for the extended HetNet model. The extended HetNet model has the same solution space as the Relaying BS model does, so we adopt a similar termination condition of the {\tt while} loop and dual variable update rule to ensure primal solution feasibility. In the {\tt while} loop, a greedy selection is applied to choose the bid that is most valued in terms of (unit price $\times$ throughput). Then the corresponding primal and dual variables are updated accordingly. However, $R_{i,k}$ in (\ref{eqn:wdp-cqi-dual}a) makes the RBs heterogeneous in the extended HetNet model. Therefore the analysis in Sec.~\ref{subsec:simple} does not carry over to the extended HetNet model in a straightforward way. 

\begin{algorithm}[!htp]
\caption{ Allocation Rules for the extended HetNet model}\label{alg:allocation2}
\begin{algorithmic}[1]
\State {// {\small Initialization}}
\State {\small $p = 0; t=0;$		$\mathcal{C}_1 = \emptyset$; $\mathcal{C}_2 = \emptyset$;}
\State {\small  $\forall i\in\mathcal{B}: x_{i} = 0$, $\tilde{\rho}_i=0$, $\tilde{\xi}_i = 0$, $\tilde{\lambda}^0_k = 1/|N|$;}
\State
\While{\small $\mathcal{C}\neq \mathcal{B}$ AND $\sum_{k}\tilde{\lambda}^t_k \leq \exp(\delta-2)$} 
	\State $(\mu,\mathcal{D}) = \arg\max_{i\in \mathcal{B}\setminus\mathcal{C}_1,\mathcal{D}\subseteq N\setminus\mathcal{C}_2:|\mathcal{D}|=r_i}\big\{\sum_{k\in \mathcal{D}}R_{i,k} \big\}$;	
	\State $x_{\mu} = 1;$ $a_{\mu,k} = 1, \forall k\in  \mathcal{D};$ 
	\State $\tilde{\xi}_{\mu} =  \sum_{k\in\mathcal{D}}R_{\mu,k}$; $p = p+\sum_{k\in\mathcal{D}}R_{\mu,k}$; 
	\State $\mathcal{C}_1 = \mathcal{C}_1\cup\{\mu\}; \mathcal{C}_2 = \mathcal{C}_2\cup\{\mathcal{D}\}$
	\If{$\mu\in \mathcal{B}_r$}
	    \State $\tilde{\rho}_{\mu} = \frac{\lambda^{t}}{|\mathcal{B}_r|}$
	\EndIf
	\State $\tilde{\lambda}^{t+1}_k = \tilde{\lambda}^{t}_k \cdot \exp(\delta-2)^{r_{\mu}/(|N|-2\max_{i\in\mathcal{B}}r_i)}$;
	\State $t = t+1;$
\EndWhile
\end{algorithmic}
\end{algorithm}

\begin{lemma}
\label{lemma:dp-runtime}
Given a set $\mathcal{E}=\{e_1,...,e_k\}$ and an integer $r \leq |\mathcal{E}|$, finding a subset $E\subseteq \mathcal{E}$ such that $E = \arg\max\{\sum_{i\in E} e_i\vert~|E| = r\}$ can be done in $O(|\mathcal{E}|r)$ time.
\end{lemma}

\noindent{\em Proof:} Define the following $f(k',r')$:
{\small
\begin{equation*}
f(k',r') = \max_{E'\subseteq\{e_1,...,e_{k'}\}:|E'|=r'}\sum_{i\in E'} e_i.
\end{equation*}}
We then have a recursive formula as follows:

{\small
\begin{equation}
\label{eqn:dp}
f(k',r') = \left\{\begin{array}{lll}
\max\big\{&f(k'-1,r'-1)+e_{k'}, &~\mbox{if}~ k' > r'>1\\
&f(k'-1, r')\big\}&\\
\sum_{i=1}^{k'} e_i &&~\mbox{if}~ k'=r'\\
0&&~\mbox{if}~ r'=0\\
\end{array}\right.
\end{equation}}

Following Eqn. \ref{eqn:dp}, we can design a dynamic programming algorithm to solve the problem. The total number of states in the dynamic programming table is $\mathcal{E}\times r$, while computing each state takes two operations. The overall time complexity is $O(\mathcal{E}\times r)$. \qed

\begin{theorem}
\label{thm:alg2_runtime}
Algorithm \ref{alg:allocation2} computes a feasible solution to the primal problem (\ref{eqn:wdp-cqi}) in $O(m|N||\mathcal{B}|^2)$ time.
\end{theorem}

\noindent {\em Proof}: The proof of the feasibility is similar to that of Theorem \ref{thm:alg1_runtime}. 
The {\tt while} loop iterates at most $|\mathcal{B}|$ times. In line 6, for each bidder $i\in\mathcal{B}\setminus\mathcal{C}_1$, we need find $\arg\max_{\mathcal{D}\subseteq N\setminus\mathcal{C}_2:|\mathcal{D}|=r_i}\big\{\sum_{k\in \mathcal{D}}R_{i,k} \big\}$, which costs at most $|N|r_i \leq |N|m$ according to Lemma \ref{lemma:dp-runtime}. The total time complexity is:  $O(|\mathcal{B}|\times |\mathcal{B}| \times |N|^2| = O(m|N||\mathcal{B}|^2)$. \qed

The next lemma shows the winning probability of a bidder increases as the bidding price $v_i$ increases in the extended HetNet model.

\begin{lemma}
\label{lemma:allocation2-monotonic}
Algorithm \ref{alg:allocation2} ensures the winning probability is monotonically non-decreasing in the bidding price.
\end{lemma}

\noindent {\em Proof}: 
The allocation rule in Algorithm \ref{alg:allocation2} is deterministic, and the winning probability is either $0$ or $1$. If bidder $i$ wins with unit bidding price $v_i$, then it wins with any unit price $v_i'>v_i$ as well because a higher $v_i$ ensures bidder $i$ is still selected in line 7. Similarly, if bidder $i$ loses with $v_i$, then it also loses with any $v_i'<v_i$. Thus there exists a value $v_i^*$ such that bidder $i$ wins with $v_i\geq v_i^*$ and loses with $v_i<v_i^*$, implying that the winning probability is non-decreasing in the bidding price.\qed

Following Theorem \ref{thm:truthful-iff}, we set the winning bidder $i$'s payment to its critical bidding price $v_i^*$, which can be obtained via binary search shown in Algorithm \ref{alg:payment1}, with the following time complexity: $O(|\mathcal{B}|\log (m/\epsilon) m|N||\mathcal{B}|^2) = O(|\mathcal{B}|^3m\log (m/\epsilon)|N|)$.  

\begin{theorem}
The auction with the allocation rules in Algorithm \ref{alg:allocation2} and payment rule shown above is truthful and individual rational.
\end{theorem}

\noindent {\em Proof}: Lemma \ref{lemma:allocation2-monotonic} and the critical bid based payments guarantee the truthfulness of the auction in the extended HetNet model.

Under truthfully bidding, the utility of a winning bidder is non-negative since the critical bidding price is always lower than the bidding price, which equals the true value. The utility of a losing bidder is zero since it receives nothing and pays zero. Thus the auction is individual rational.
\qed

\section{Evaluation}
\label{sec:experiment}

\subsection{Simulation Studies}
\begin{figure}[!htbp]
	\begin {center}
    \begin{minipage}[t]{0.49\linewidth}
    \centering
	\includegraphics[width=\textwidth]{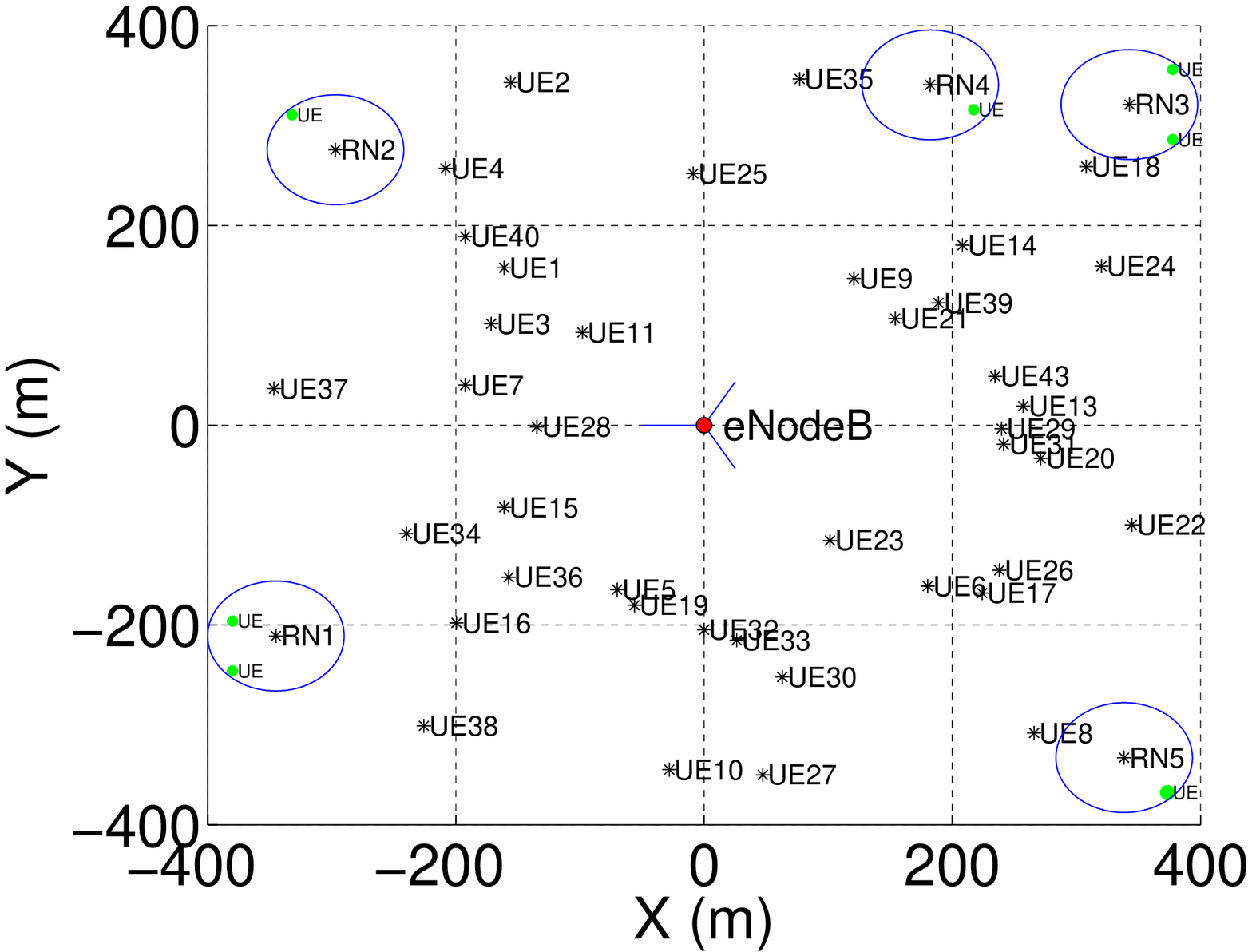}
	\caption{Initial positions of eNodeB, UEs and RNs.}
	\label{fig:position}
    \end{minipage}
    \hfill
    \begin{minipage}[t]{0.49\linewidth}
    \centering
	\includegraphics[width=\textwidth]{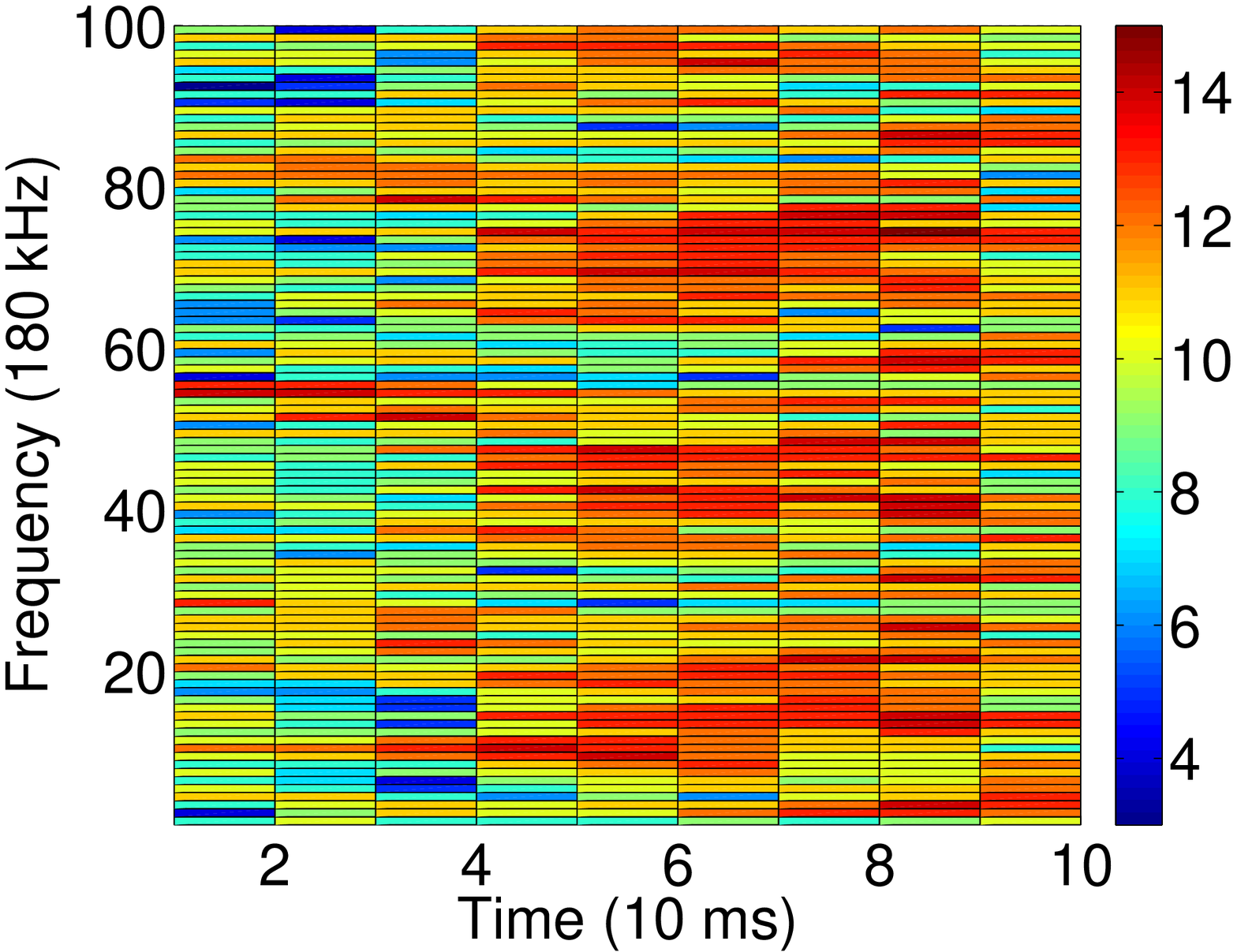}
	\caption{UE5'CQI varies over time and frequency.}
	\label{fig:cqi}
    \end{minipage}
	\end {center}
\end{figure}

To verify the performance of the proposed mechanisms, we conduct a large-scale simulation studies. An LTE simulator \cite{VTC2010} is employed to generate the SINR as well as corresponding CQIs according to the TS25.814 pathloss model. The generated traces are used to drive large-scale simulation studies to verify the auction performance. The system runs in 900 MHz with a bandwidth of 20 MHz. There are 47 UEs and 5 RNs in one donor {\em eNodeB}'s coverage. 40 out of 47 UEs are served by the donor {\em eNodeB} directly. The other $7$ UEs are located in the coverage of RNs and served by the RNs instead. The positions of the {\em eNodeB}, UEs and RNs are illustrated in Figure \ref{fig:position}. All UEs are walking with a speed of 5 km/h in given directions. The donor {\em eNodeB} has a transmit power of 49 dBm. The entire simulation spans 10 $s$ where each time slot lasts 10 $ms$, {\em i.e.}, 1000 time slots are simulated. Each user's resource block demand is uniformly distributed among $[10,40]$ since the demand pattern can be considered as uniformly random when the duration is rather short, {\em e.g.}, 10 $s$. The bidding price is generated according to standard pricing with a major LTE Advanced carrier in North America, {\em e.g.}, \$15 for 300 MB, with randomness. The simulation setting is summarized in Tab. \ref{tab:experiment}. Figure \ref{fig:cqi} illustrates a generated trace, {\em i.e.}, UE5's CQIs over time and frequency.

\begin{table}[!htbp]
\caption{Evaluation Settings}
\centering
\label{tab:experiment}
{\small
\begin{tabular}{|c|c|c|c|}
\hline
\hline
\# of {\em DeNB}s  &1 &Frequency & 900 MHz\\ 
\hline 
\# of RNs &5 &Bandwidth & 20 MHz\\
\hline
\# of UEs served & & {\em eNodeB} TX power & 49 dBm\\\cline{3-4}
 by {\em DeNB}&  40&UEs' speed & 5 km/h\\
\hline
\# of UEs served & &UEs' position & Figure \ref{fig:position}\\\cline{3-4}
by each RN& 1 or 2& Auction frequency & every 10 ms\\
\hline
\end{tabular}}
\end{table}

\subsubsection{The Relaying Base Station Model}
\label{subsec:simple}

\begin{figure}[htbp]
	\begin {center}
	\begin{minipage}[t]{0.48\linewidth}
    \centering
    \includegraphics[width=\textwidth]{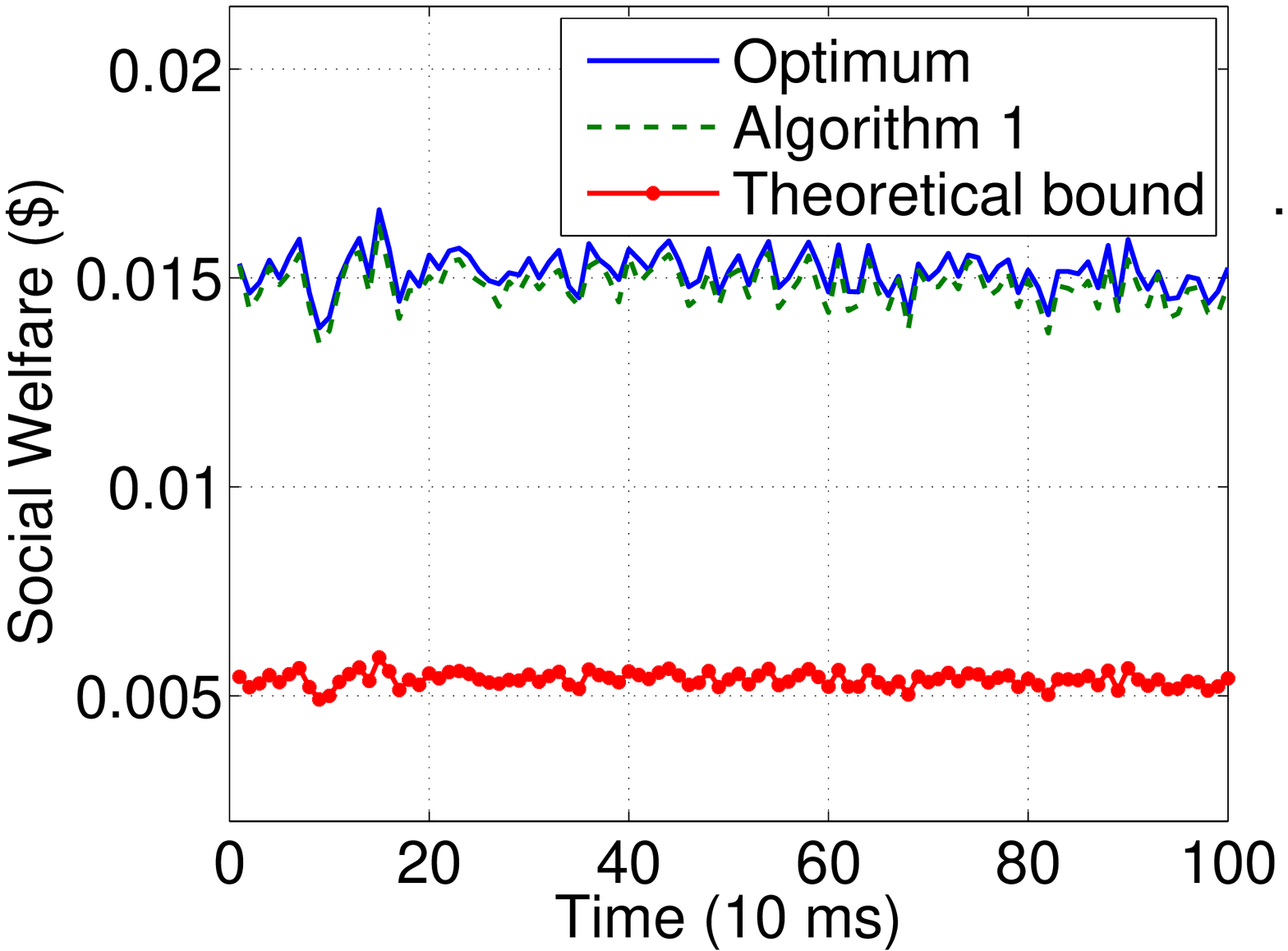}
	\caption{Social welfare in Algorithm \ref{alg:greed} compared with the Optimum for the first 1000 $ms$.}
	\label{fig:alg1-socialwelfare1}
    \end{minipage}
    \hfill
    \begin{minipage}[t]{0.48\linewidth}
    \centering
    \includegraphics[width=\textwidth]{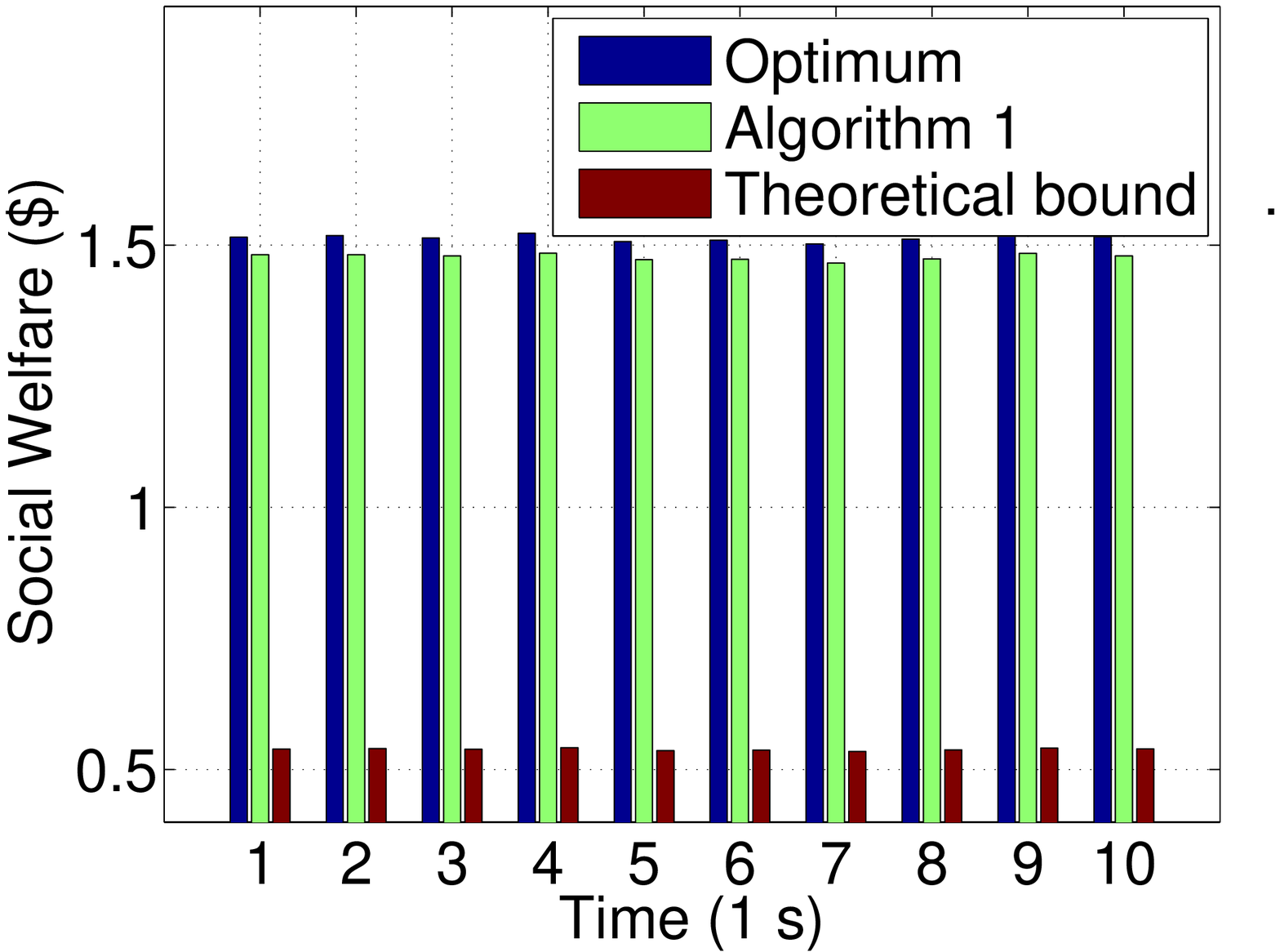}
    \caption{Social welfare Comparison: Algorithm \ref{alg:greed} and the Optimum, for $t = 1\sim 10 s$.}
    \label{fig:alg1-socialwelfare2}
    \end{minipage}
   	\end {center}
\end{figure}

\noindent {\bf Social Welfare}. Algorithm \ref{alg:greed} is run for the given simulation setting. We first compare the social welfare obtained by Algorithm \ref{alg:greed} with the optimal one as well as the lower bound of the social welfare computed according to Theorem \ref{thm:alg1-apprx-ratio}, as shown in Figure \ref{fig:alg1-socialwelfare1}, for the first 1000 $ms$. We employ {\tt cplex} as the integer programming solver to solve the problem optimally. We observe that the social welfare of Algorithm \ref{alg:greed} is rather close to the optimum, and is substantially better than the proven theoretical bound. The ratio in Theorem \ref{thm:alg1-apprx-ratio} is for the worst case, and can be rather loose in many cases in practices. The same conclusion is true for the entire simulation duration, as illustrated in Figure \ref{fig:alg1-socialwelfare2}.

\begin{figure}[htbp]
	\begin {center}
    \begin{minipage}[t]{0.48\linewidth}
    \centering
    \includegraphics[width=\textwidth]{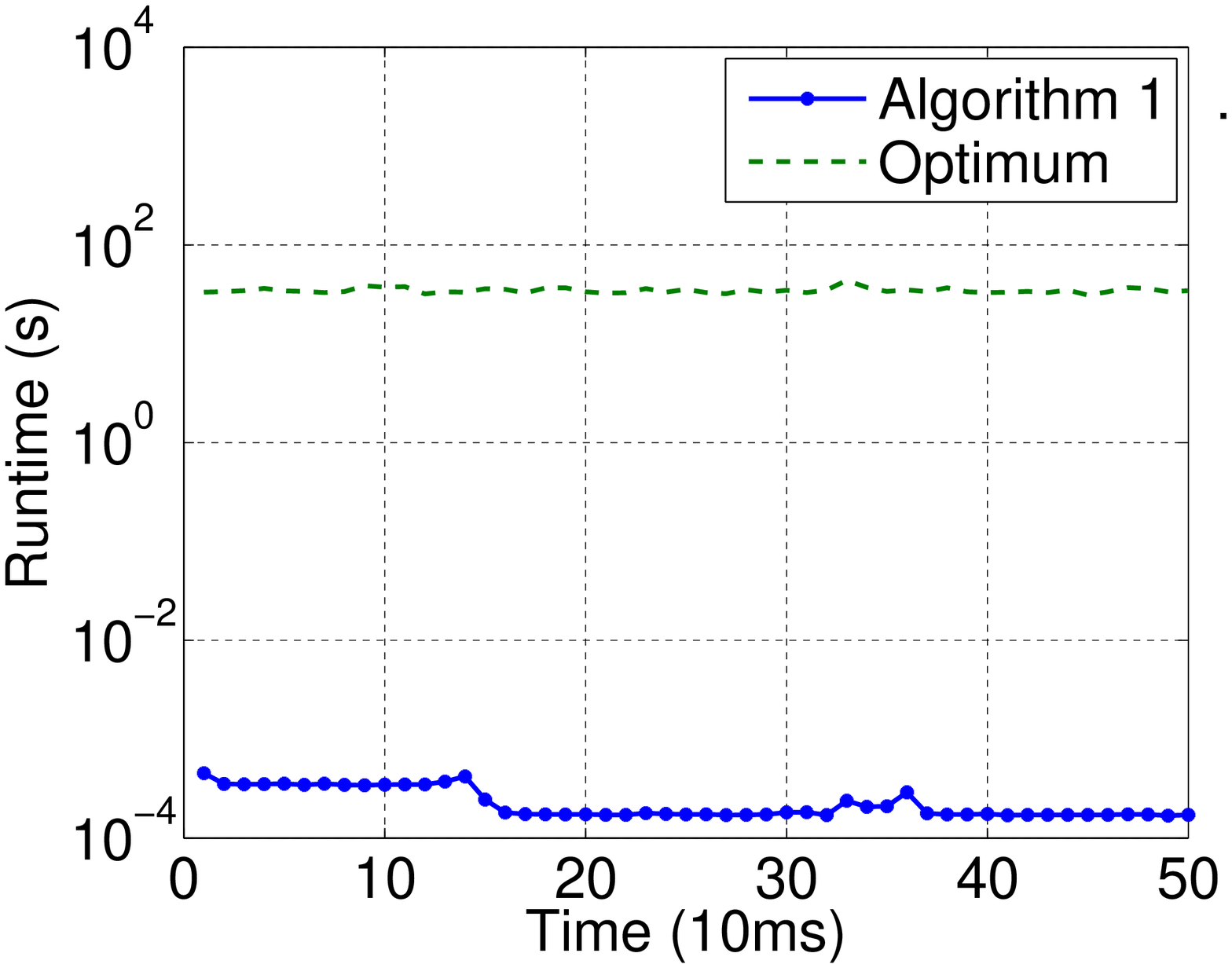}
	\caption{Runtime comparison between Algorithm \ref{alg:greed} and the Optimum for the first 500 $ms$.}
	\label{fig:alg1-runtime}
    \end{minipage}
    \hfill
    \begin{minipage}[t]{0.48\linewidth}
    \centering
    \includegraphics[width=\textwidth]{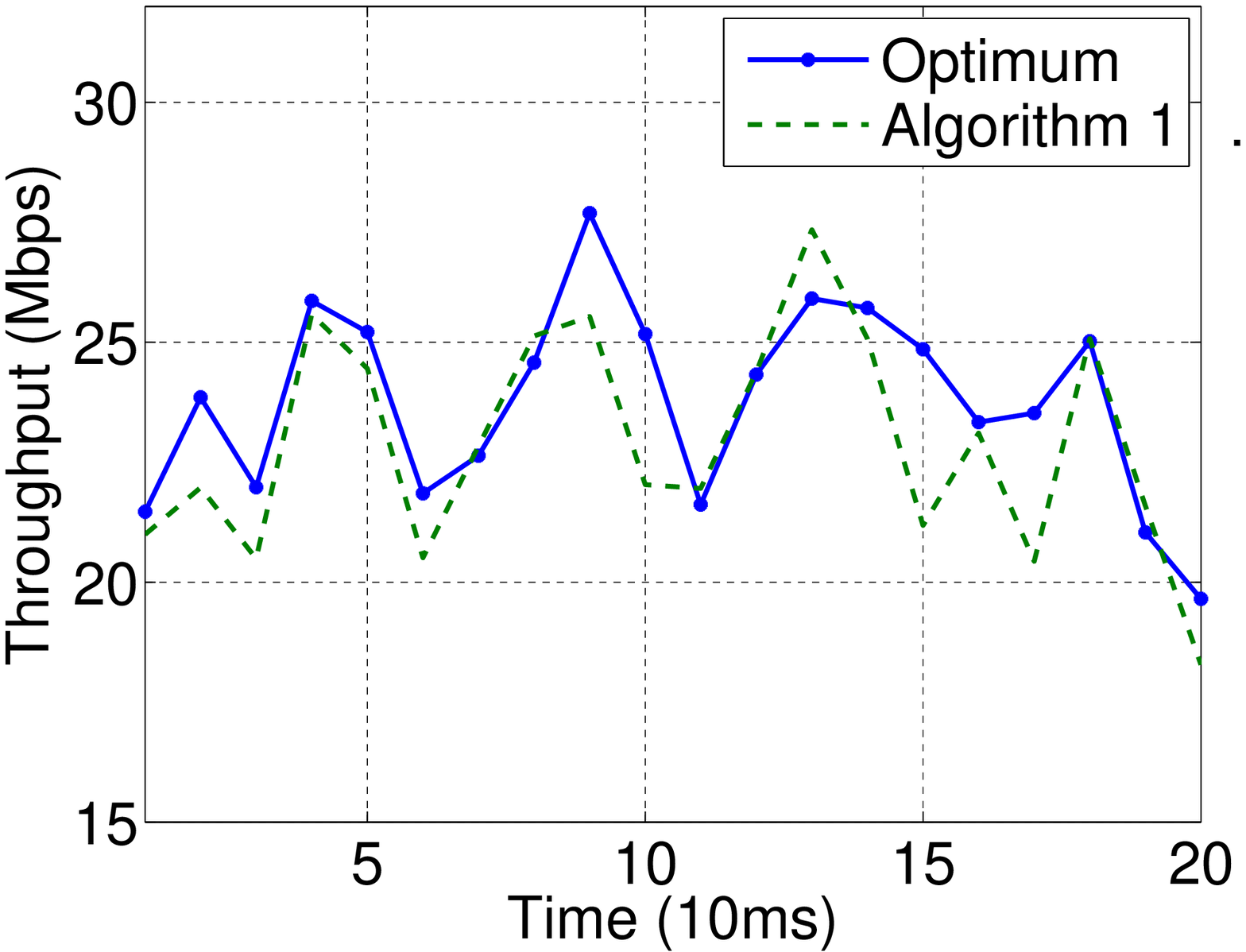}
    \caption{Throughput comparison: Algorithm \ref{alg:greed} and the Optimum, for $t = 1\sim 200 ms$.}
    \label{fig:alg1-throughput1}
    \end{minipage}
  	\end {center}
\end{figure}

\noindent {\bf Runtime \& Throughput.} The optimum can be computed but at costs of considerable time and computing resources. We compare the runtime of Algorithm \ref{alg:greed} along with its corresponding payment calculation rules with those of the optimal allocation. Figure \ref{fig:alg1-runtime} depicts the runtime with a logarithmic scale on time axis. We observe that the optimal algorithm is rather costly, approximately 30 $s$ per turn. It is computationally infeasible to execute in practice.
In comparison, the auction with Algorithm \ref{alg:greed} and Algorithm \ref{alg:payment1} takes much shorter time at around 0.2 $ms$ per turn. The performance can be further improved by optimizing the program structure and code. Figure \ref{fig:alg1-throughput1} illustrates the throughput of the donor {\em eNodeB} over time. We note that the optimal algorithm usually achieves higher throughput than Algorithm \ref{alg:greed} does. However, Algorithm \ref{alg:greed} works better at $t=80 ms, 110 ms, 130 ms$. Because the optimal algorithm focuses on maximizing the social welfare rather than the throughput, and other algorithms could achieve higher throughput than it.

\begin{figure}[htbp]
	\begin {center}
	\begin{minipage}[t]{0.48\linewidth}
    \centering
    \includegraphics[width=\textwidth]{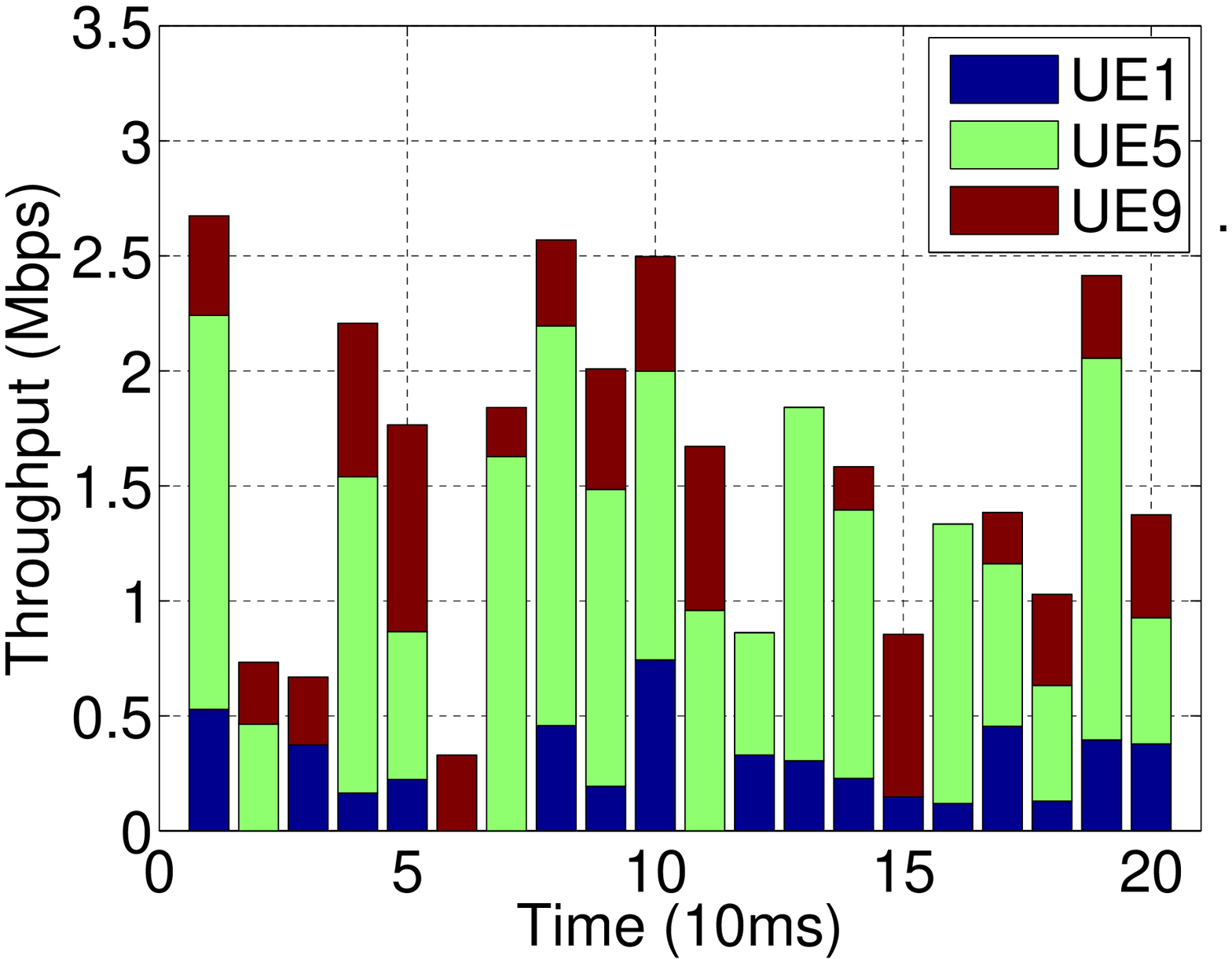}
    \caption{UEs' throughput comparison for $t = 1\sim 200 ms$.}
    \label{fig:alg1-throughput2}
    \end{minipage}
    \hfill
    \begin{minipage}[t]{0.48\linewidth}
    \centering
    \includegraphics[width=\textwidth]{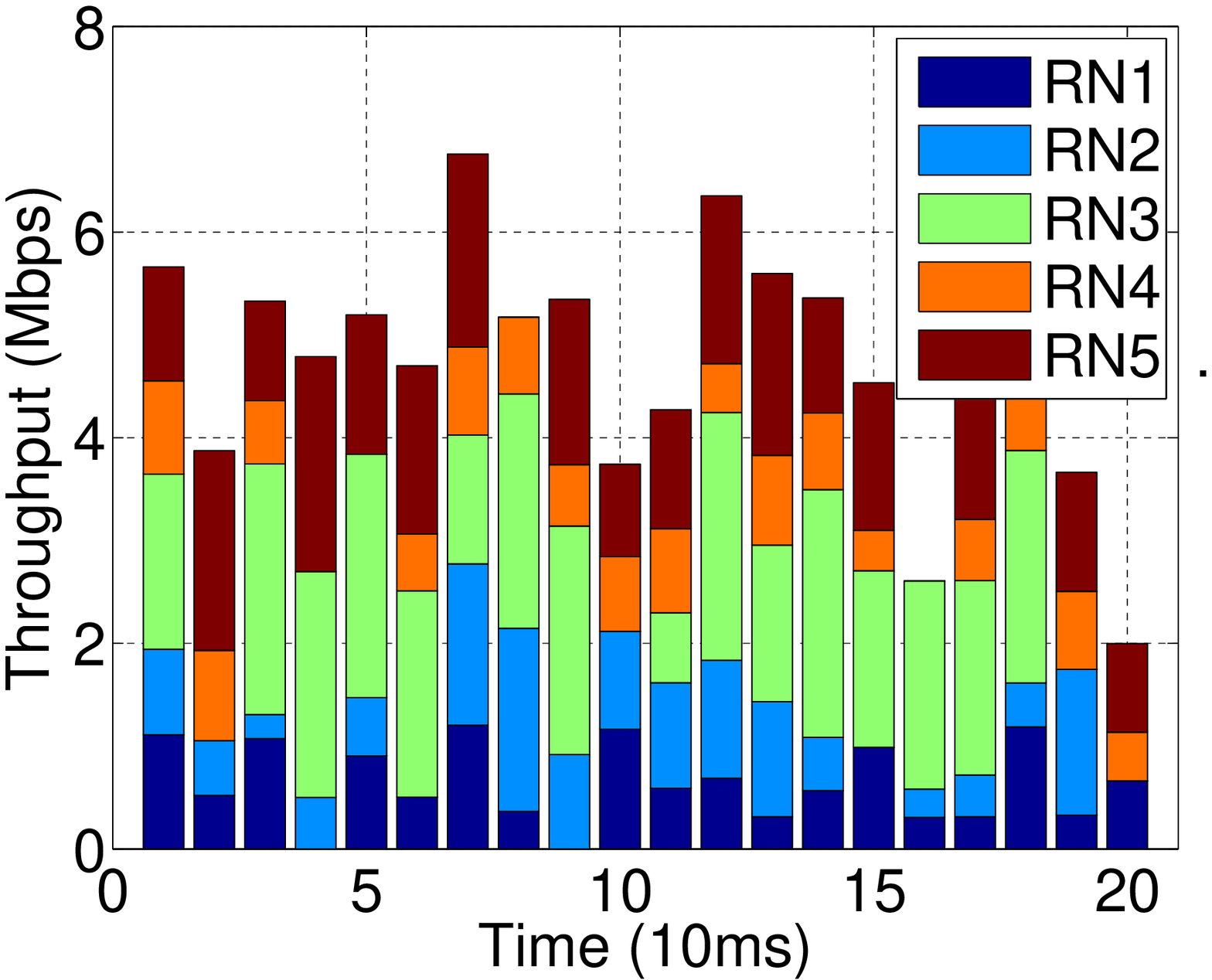}
    \caption{RNs' throughput comparison for $t=1\sim 200 ms$.}
    \label{fig:alg1-throughput3}
    \end{minipage}
   	\end {center}
\end{figure}

We also compare the throughput among UEs and RNs as shown in Figure \ref{fig:alg1-throughput2} and Figure \ref{fig:alg1-throughput3}. We observe that UEs in Figure \ref{fig:alg1-throughput2} receive sufficient downlink resource to achieve acceptable throughput. RNs also experience fair throughput, which implies that the UEs served by RNs also enjoy good throughput and the coverage is improved.

\begin{figure}[htbp]
	\begin {center}
	\begin{minipage}[t]{0.48\linewidth}
    \centering
    \includegraphics[width=\textwidth]{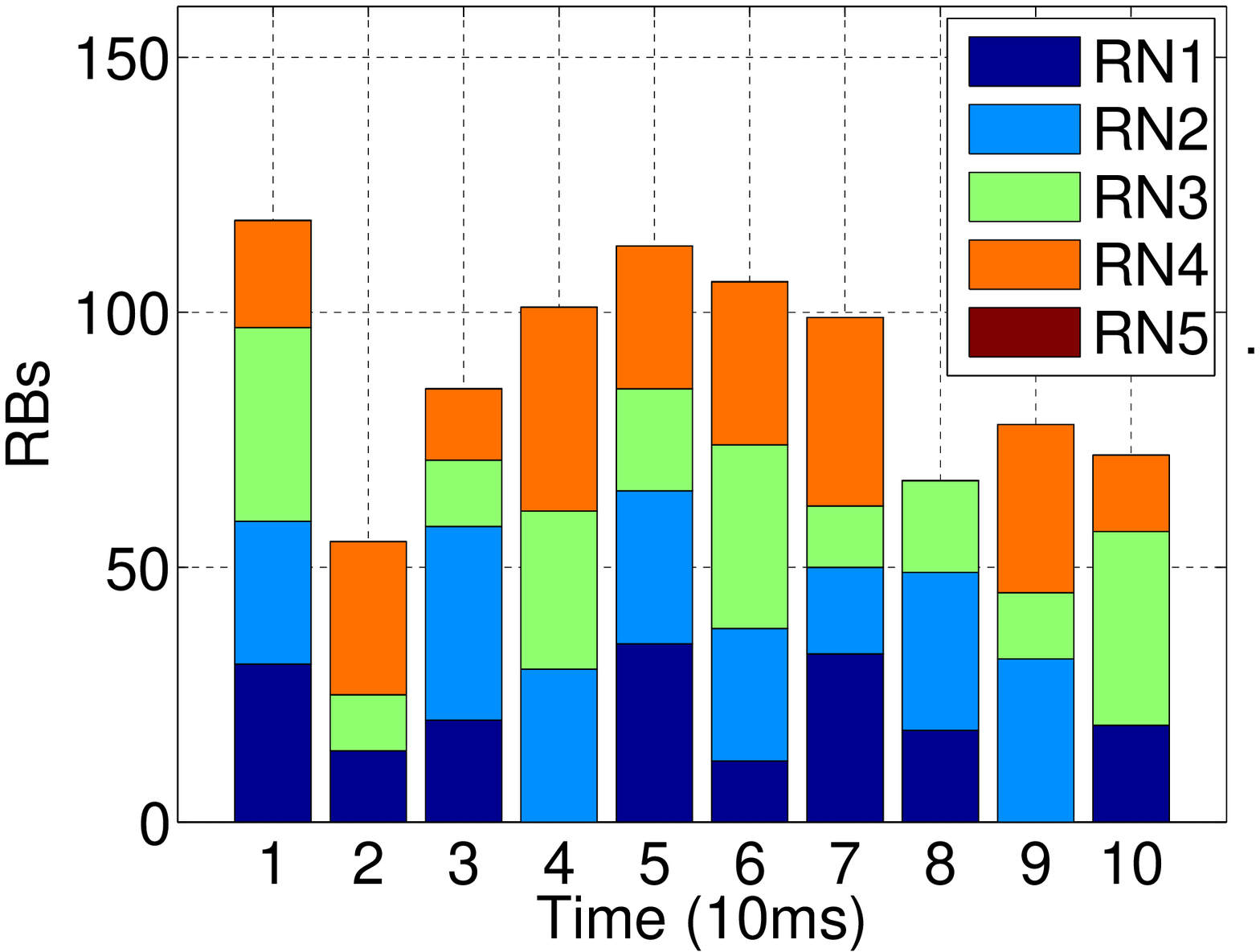}
    \caption{The \# of RBs received by RNs for $t=1\sim 200 ms$.}
    \label{fig:alg1-allocated-rb1}
    \end{minipage}
    \hfill
    \begin{minipage}[t]{0.48\linewidth}
    \includegraphics[width=\textwidth]{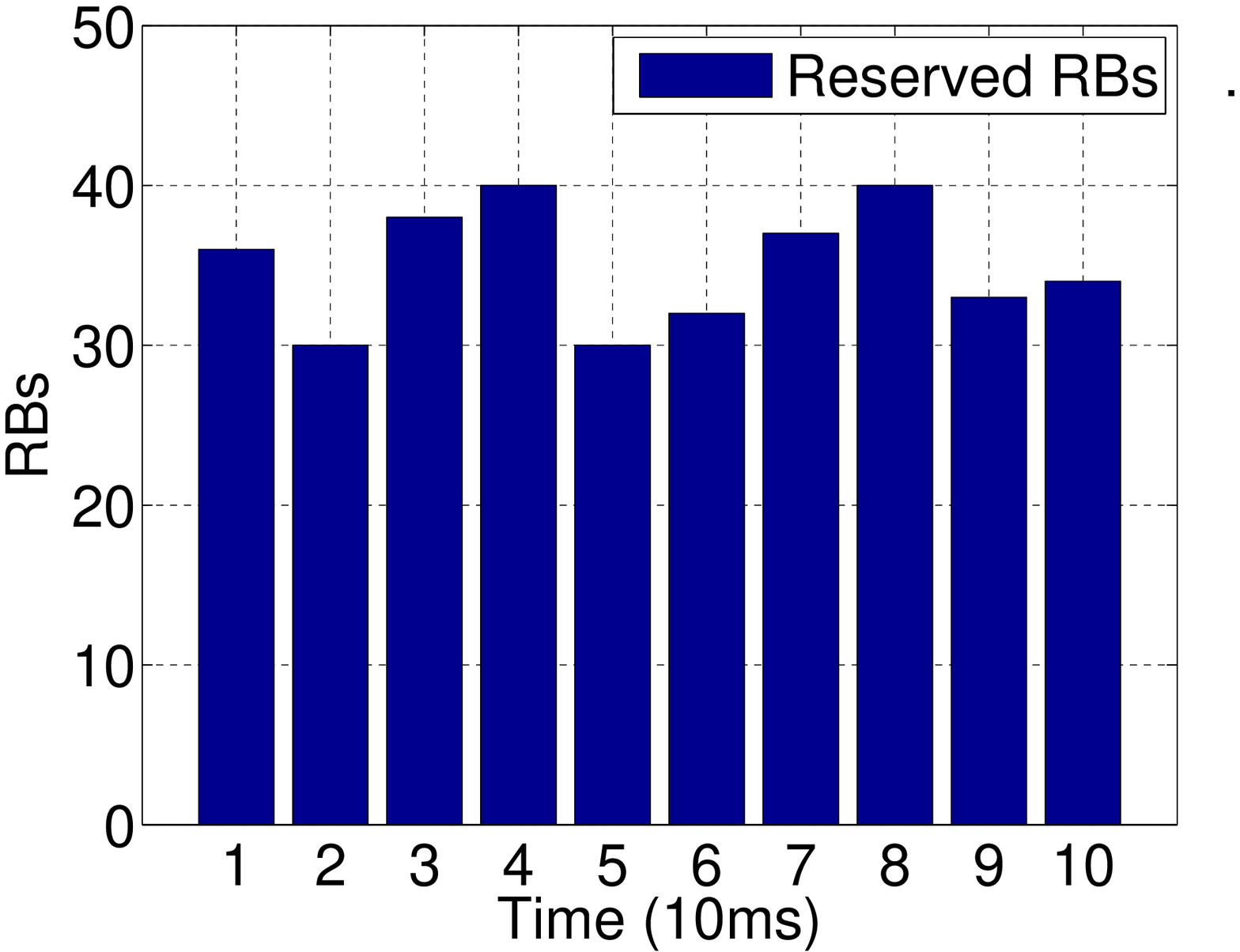}
    \caption{The \# of RBs reserved for RNs inner communication.}
    \label{fig:rn-extra-rb}
    \end{minipage}
	\end {center}
\end{figure}

\noindent {\bf RNs' Received Resources.} 
We next investigate resources allocated to the RNs. We collect the simulation trace for the first 200 $ms$, illustrate the resource block allocation in Figure \ref{fig:alg1-allocated-rb1}. In order to avoid interference with donor {\em eNodeB}, RNs are allocated extra resource blocks which are used for inner communication, as shown in Figure \ref{fig:rn-extra-rb}. The number of allocated resource blocks is more than all numbers of resource blocks received by RNs, therefore the inner communication can be guaranteed interference-free.

\subsubsection{Extended HetNet Model}
We employ the same UEs and CQI traces to drive the simulation for the extended HetNet model. The bids then are regenerated according to the requirement of the extended HetNet model and therefore different from those in Sec. \ref{subsec:simple}.

\begin{figure}[htbp]
	\begin {center}
	\begin{minipage}[t]{0.48\linewidth}
    \centering
    \includegraphics[width=\textwidth]{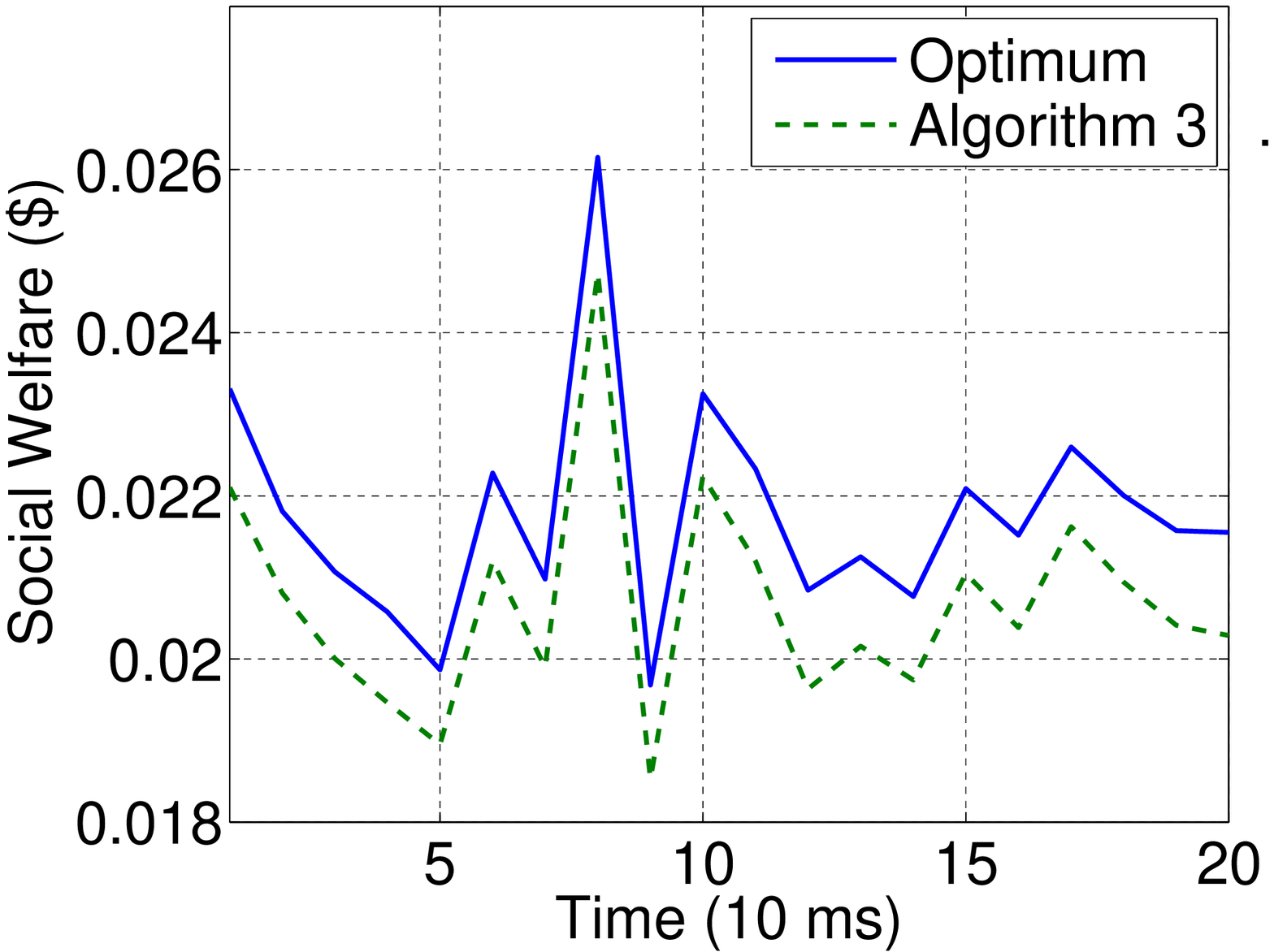}
    \caption{Social welfare comparison: Algorithm \ref{alg:allocation2} and the Optimum, for $t = 1\sim 200 ms$.}
    \label{fig:alg2-social-welfare1}
    \end{minipage}
    \hfill
    \begin{minipage}[t]{0.48\linewidth}
    \centering
    \includegraphics[width=\textwidth]{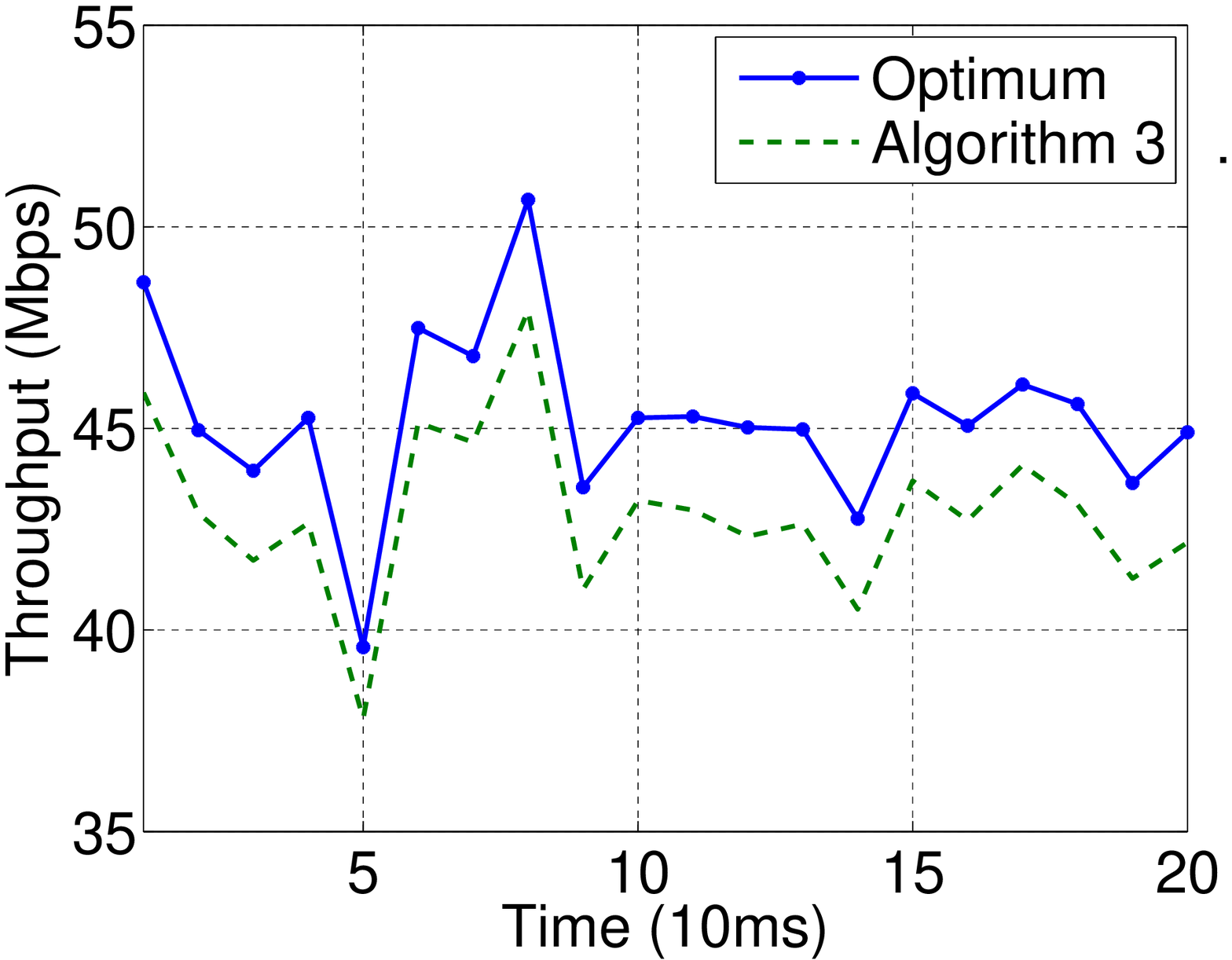}
    \caption{Throughput comparison: Algorithm \ref{alg:allocation2} and the Optimum, for $t = 1\sim 200 ms$.}
    \label{fig:alg2-throughput1}
    \end{minipage}
   	\end {center}
\end{figure}

\noindent {\bf Social Welfare.} The proposed allocation algorithm achieves close-to-optimum performance in terms of social welfare for the given problem instance, as illustrated in Figure \ref{fig:alg2-social-welfare1}. The observed gap between Algorithm \ref{alg:allocation2} and the optimum is less than 6\%. We also compute the ratios for first ten time slots as shown in Table \ref{tab:ratio-alg2}, and the average ratio is $0.95$.

\begin{table}[!htbp]
\caption{Approximation Ratio of Algorithm \ref{alg:allocation2}}
\centering
\label{tab:ratio-alg2}
{\small
\begin{tabular}{|c|c|c|c|c|c|}
\hline
t: 1 $\sim$ 5& 0.948 & 0.954 & 0.950 & 0.946 &0.954 \\
\hline
t: 6 $\sim$ 10& 0.952 & 0.949 & 0.946 & 0.942 & 0.956\\
\hline
\end{tabular}}
\end{table}

\begin{figure}[htbp]
	\begin {center}
	\begin{minipage}[t]{0.48\linewidth}
    \centering
    \includegraphics[width=\textwidth]{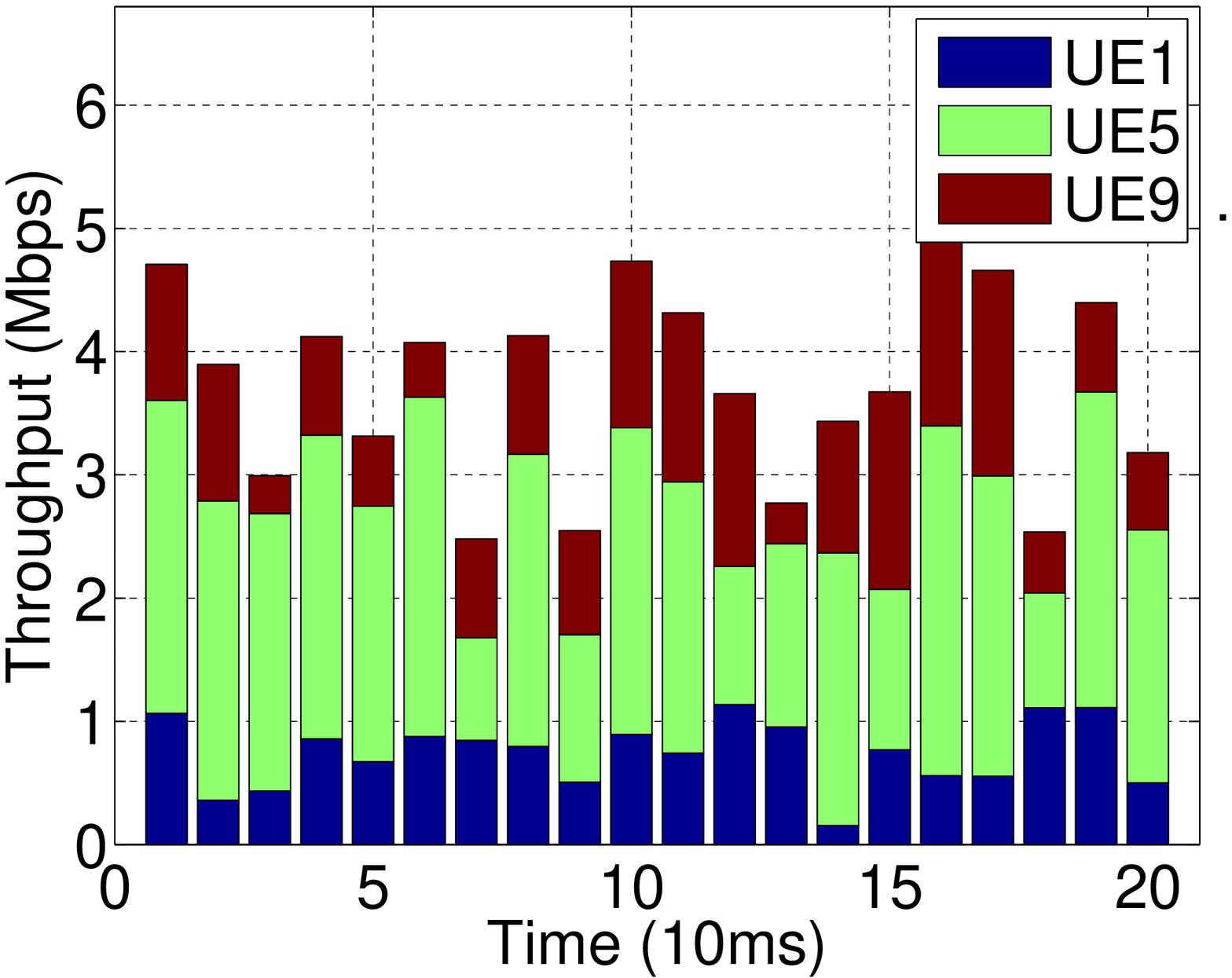}
    \caption{UEs' throughput comparison for $t = 1\sim 200 ms$.}
    \label{fig:alg2-throughput2}
    \end{minipage}
    \hfill
    \begin{minipage}[t]{0.48\linewidth}
    \centering
    \includegraphics[width=\textwidth]{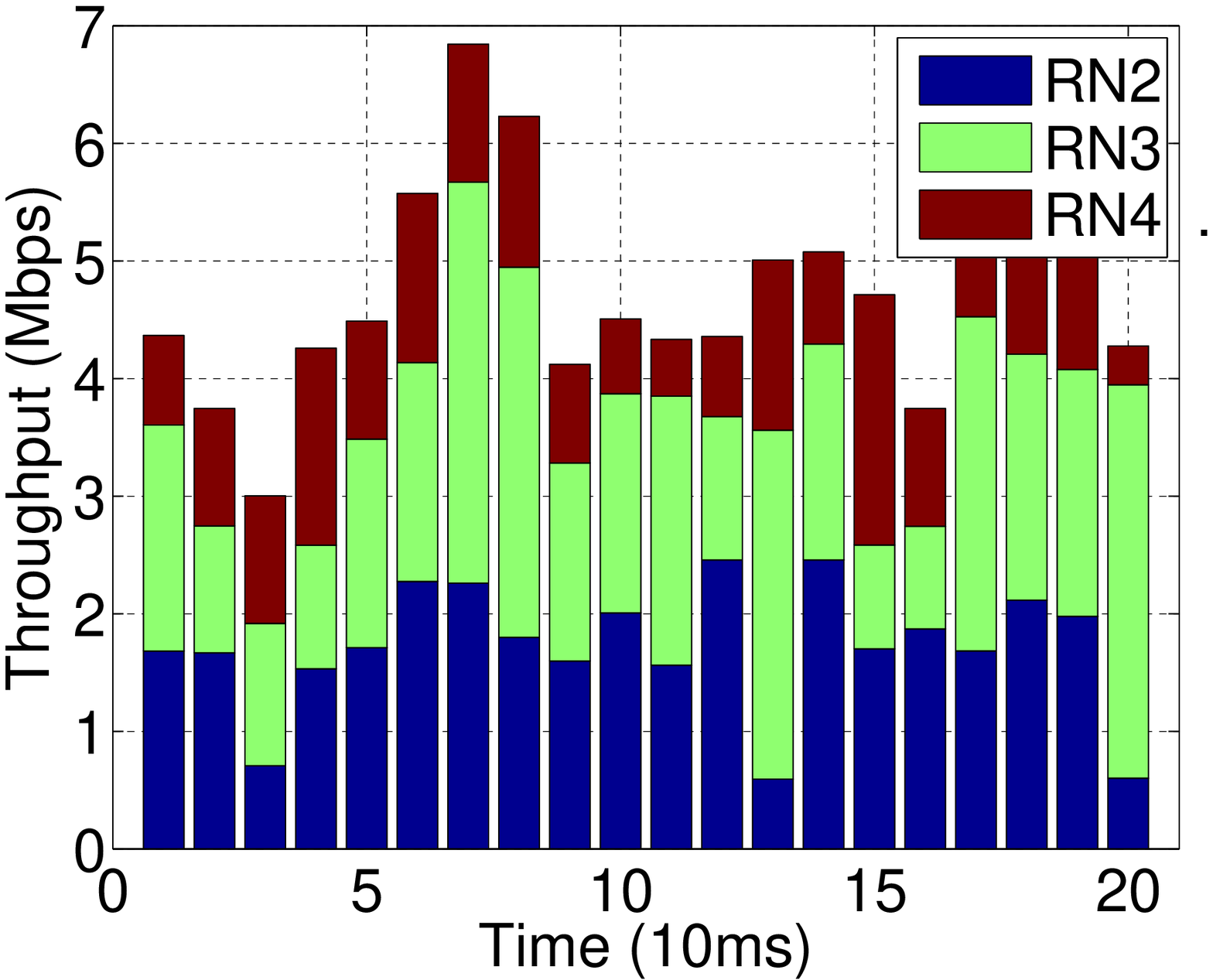}
    \caption{RN's throughput comparison for $t = 1\sim 200 ms$.}
    \label{fig:alg2-throughput3}
    \end{minipage}
   	\end {center}
\end{figure}

\noindent {\bf Throughput.} We next investigate the performance of Algorithm \ref{alg:allocation2} in terms of throughput. Figure \ref{fig:alg2-throughput1} illustrates the throughput curves of Algorithm \ref{alg:allocation2} and the optimum over $t = 1\sim 200 ms$. Different from the throughput comparison in Sec.~\ref{subsec:simple}, the optimum always achieves higher throughput than Algorithm \ref{alg:allocation2} does. Furthermore, the extended HetNet model achieves higher total throughput than the relaying base station model, {\em e.g.}, the average throughput of Algorithm \ref{alg:greed} in Figure \ref{fig:alg1-throughput1} is 22.9 $Mbps$ while the average throughput of Algorithm \ref{alg:allocation2} is 42.9 $Mbps$.  This is because the extended HetNet model takes CQI into account for opportunistic assignment. A UE with higher CQI has a higher chance to receive resources, {\em i.e.}, experiencing high throughput. Compared with the relaying base station model, each individual UE also has higher throughput, as shown in Figure \ref{fig:alg2-throughput2}. Similar conclusion holds for RNs as well, as illustrated in Figure \ref{fig:alg2-throughput3}.

\begin{figure}[htbp]
	\begin {center}
    \subfigure[Relaying Base Station Model]{
	\begin{minipage}[t]{0.48\linewidth}
    \centering
    \includegraphics[width=\textwidth]{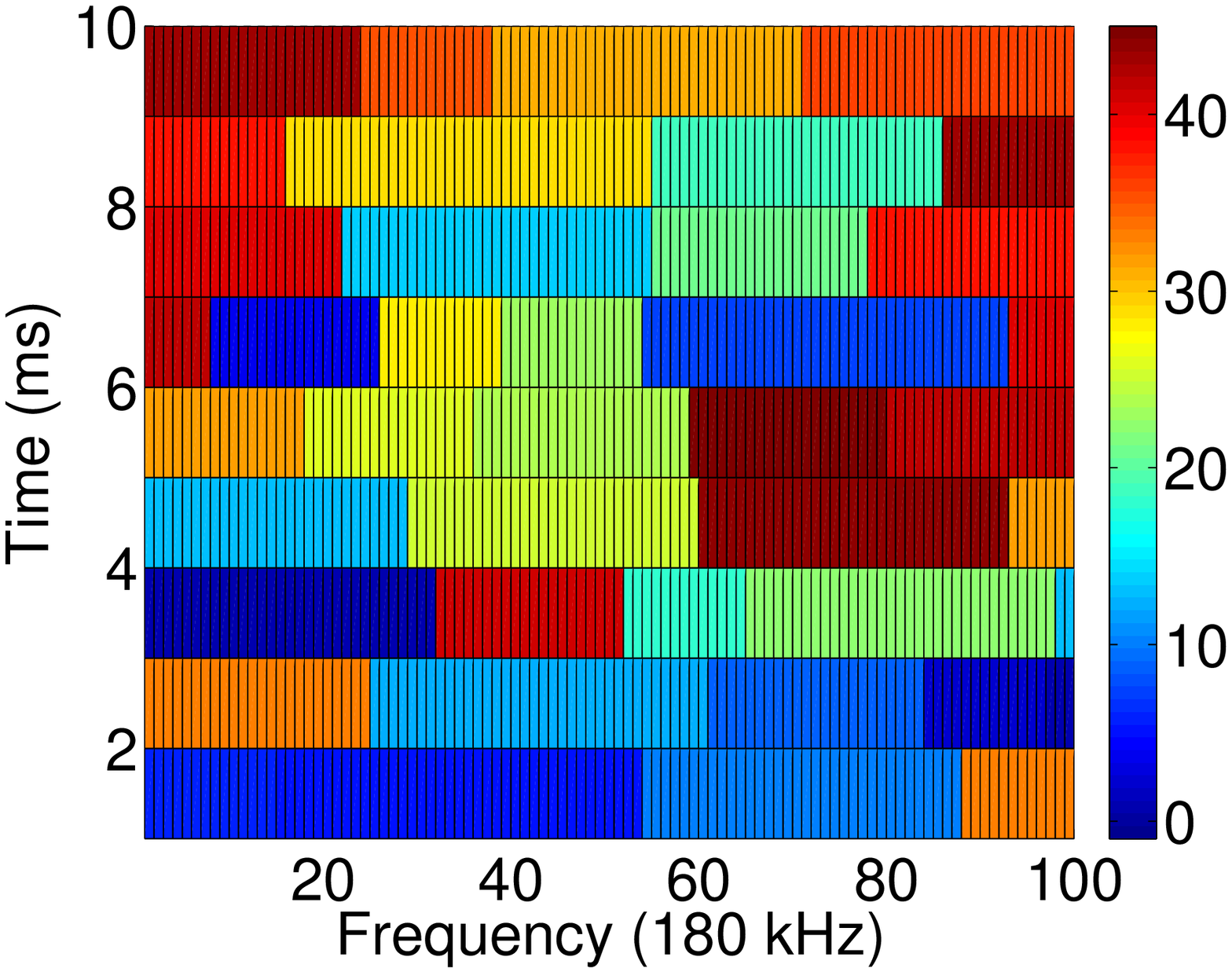}
    \end{minipage}}
    \hfill
    \subfigure[Extended HetNet Model]{
    \begin{minipage}[t]{0.48\linewidth}
    \centering
    \includegraphics[width=\textwidth]{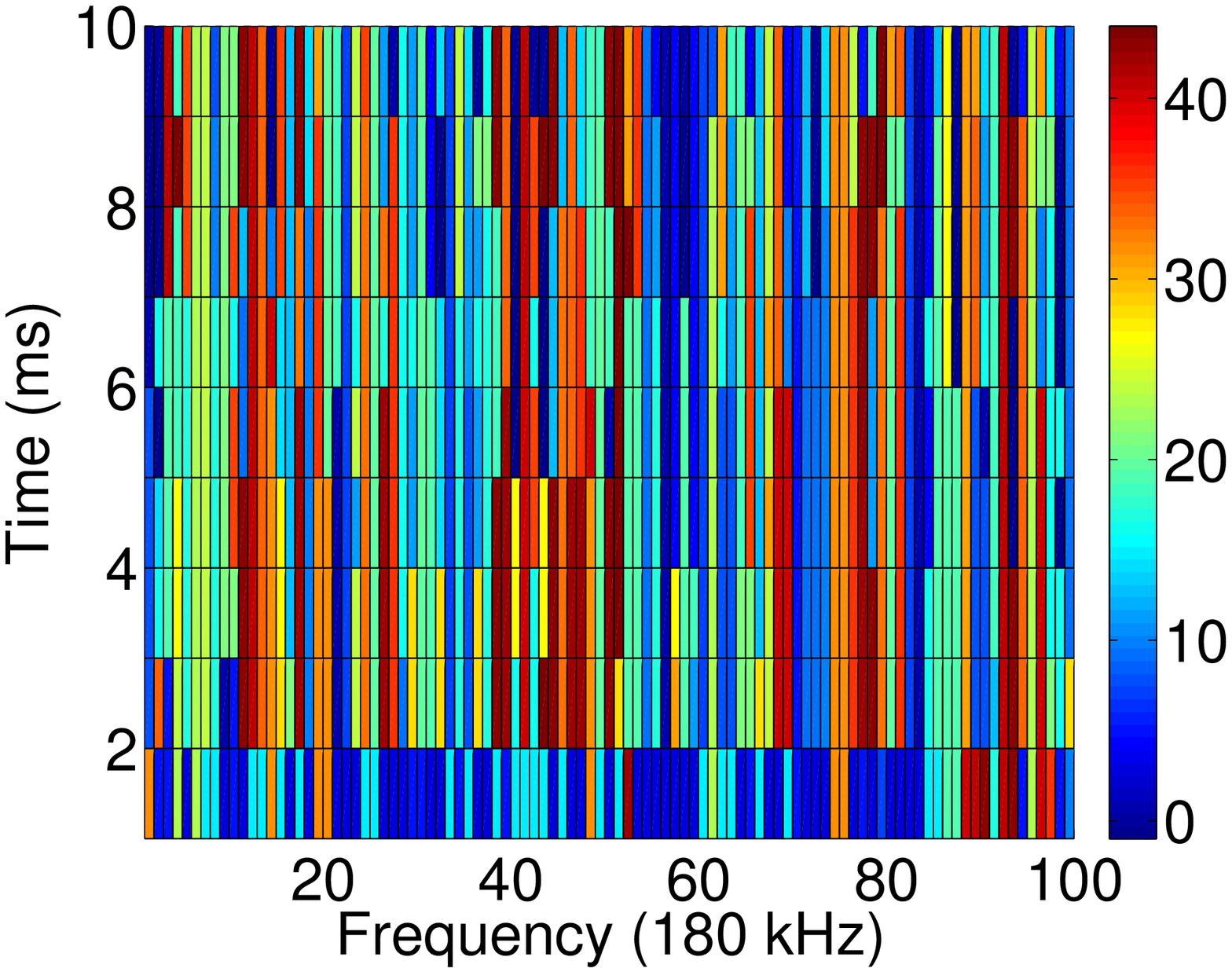}
    \end{minipage}}
   	\end {center}
    \label{fig:allocation1}
    
    \caption{RB allocation results in 1st turn.}
\end{figure}

\noindent {\bf Resource Allocation Results.} We next study the allocation result. We compare two results from the relaying base station model and the extended HetNet model, respectively, as shown in Figure \ref{fig:allocation1}. Each UE or RN is represented by a color. We notice that the relaying base station model intends to allocate consecutive resource blocks to a user while the extended HetNet model would select the resource blocks that can transmit more bits.

\noindent {\bf Comparison with Other Schedulers.} Our proposed mechanism is compared with two existing schedulers, the Round Robin scheduler and the Best CQI scheduler \cite{lte-sim-pl}, in terms of social welfare and throughput. The Best CQI scheduler assigns resource blocks to the user who has best CQI value, {\em i.e.}, best radio link condition. While Round Robin allocates resource blocks to users in turn, {\em i.e.}, all users receive equal size of resource blocks over time. 

\begin{figure}[htbp]
	\begin {center}
    \begin{minipage}[t]{0.48\linewidth}
    \centering
    \includegraphics[width=\textwidth]{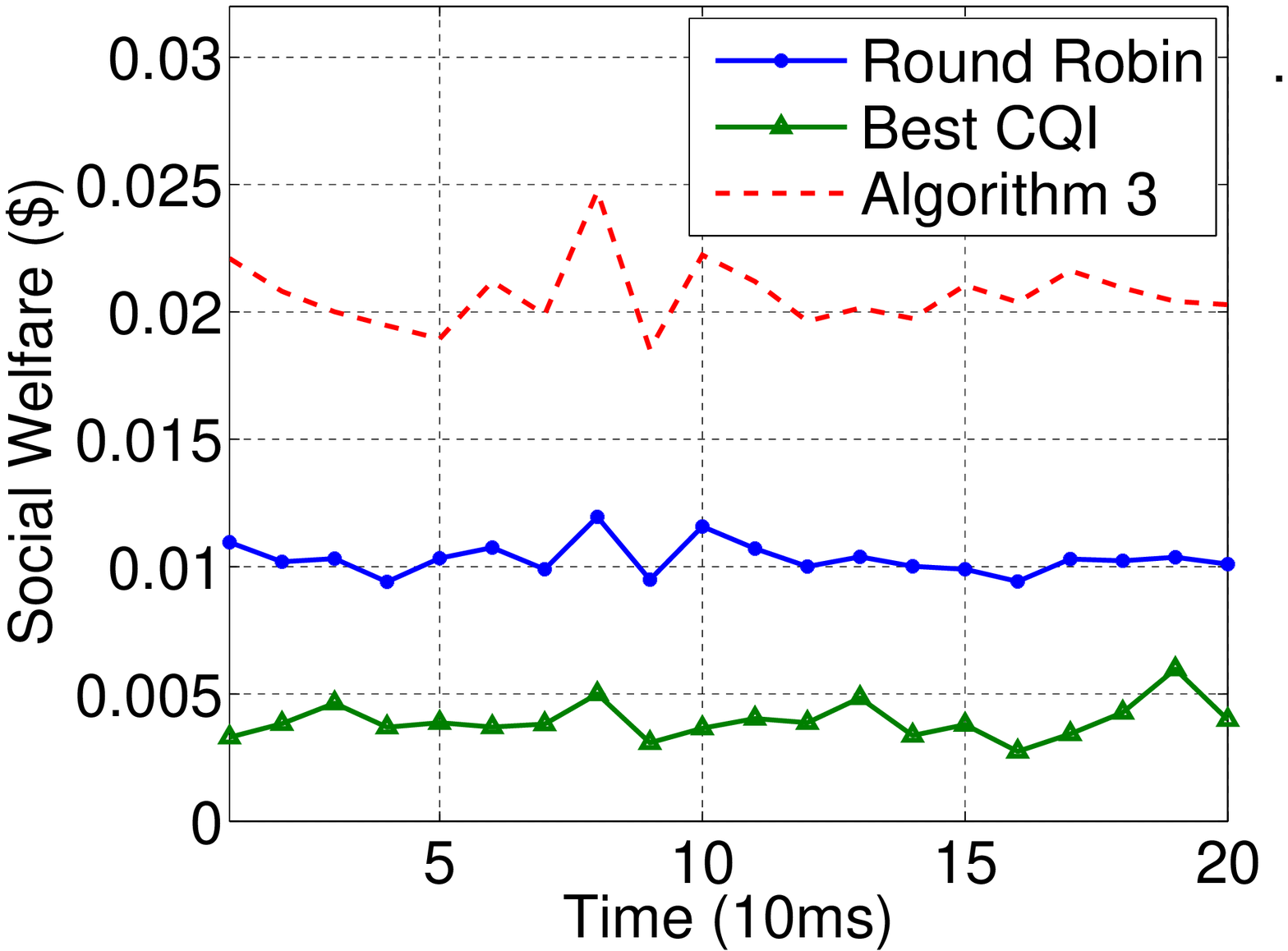}
    \caption{Social welfare comparison for $t= 1\sim 200 ms$.}
    \label{fig:rrbest-social1}    
    \end{minipage}
    \hfill
	\begin{minipage}[t]{0.48\linewidth}
    \centering
    \includegraphics[width=\textwidth]{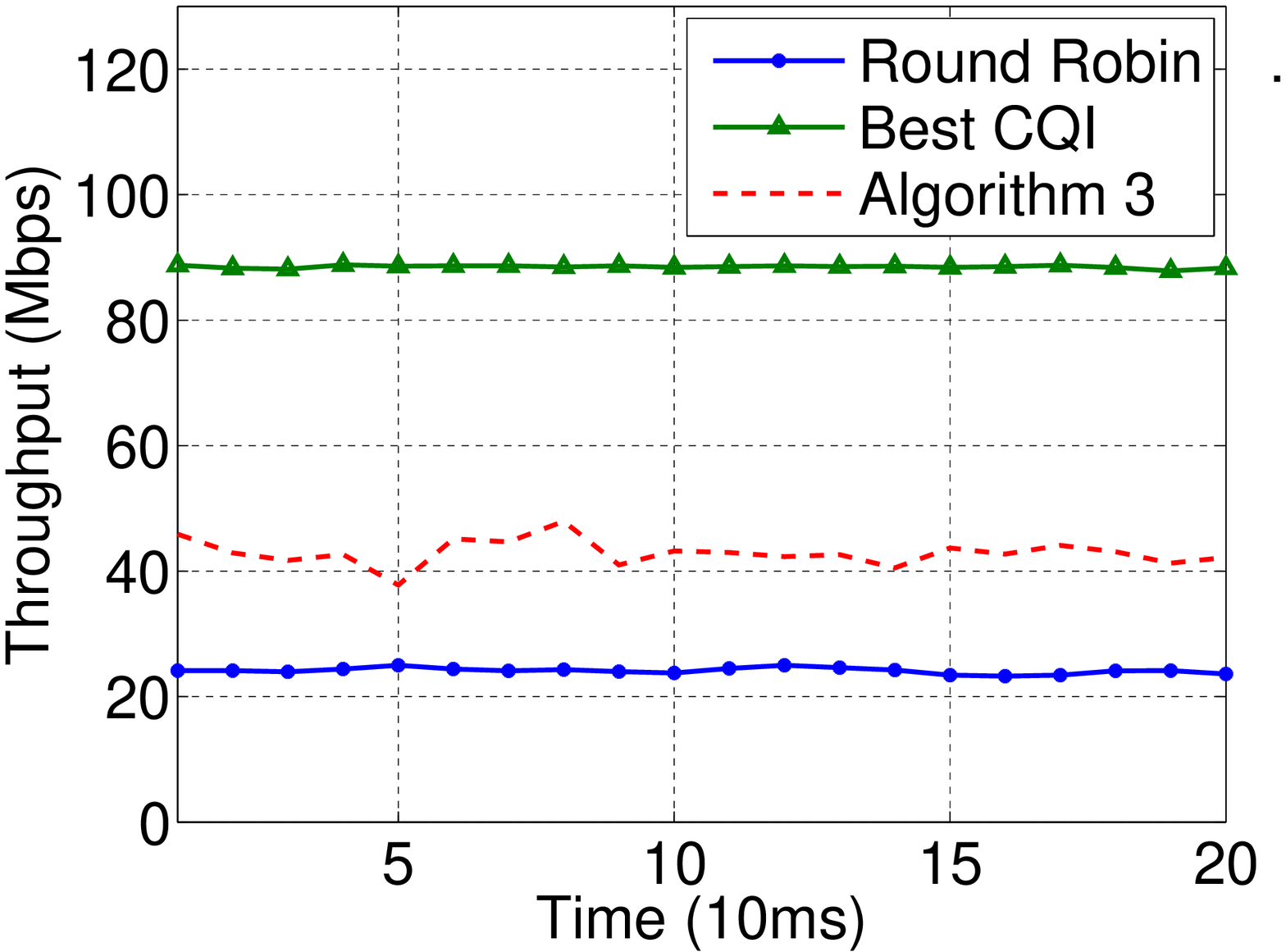}
    \caption{Total throughput comparison for $t = 1\sim 200 ms$.}
    \label{fig:rrbest-throughput1}
    \end{minipage}
   	\end {center}
\end{figure}

Figure \ref{fig:rrbest-social1} depicts the social welfare of each scheduler. We observe that Algorithm \ref{alg:allocation2} produces the highest social welfare for each time slot among these three schedulers. Round Robin is the second highest while Best CQI has the lowest social welfare. This is because both Round Robin and Best CQI allocate resources to users no matter whether they need the resources or not. Our proposed auction algorithm, on the contrary, finds how users value the resources and how many resource blocks they need from the bidding information.

Figure \ref{fig:rrbest-throughput1} shows the total throughput of each scheduler. We see that the Best CQI scheduler achieves the highest throughput, our proposed algorithm has median throughput and Round Robin experiences the lowest throughput. The reason is that: the Best CQI scheduler focuses on finding the user with the best CQI, therefore it could achieve the maximum throughput. But it fails to consider users' needs, {\em e.g.}, users with low priority traffic still receive high transmission rates. Our Algorithm \ref{alg:allocation2} tries to allocate resources to where they are valued most. It achieves the highest weighted throughput but not the highest unweighted throughput. 

\begin{figure}[htbp]
	\begin {center}
	\begin{minipage}[t]{0.48\linewidth}
    \centering
    \includegraphics[width=\textwidth]{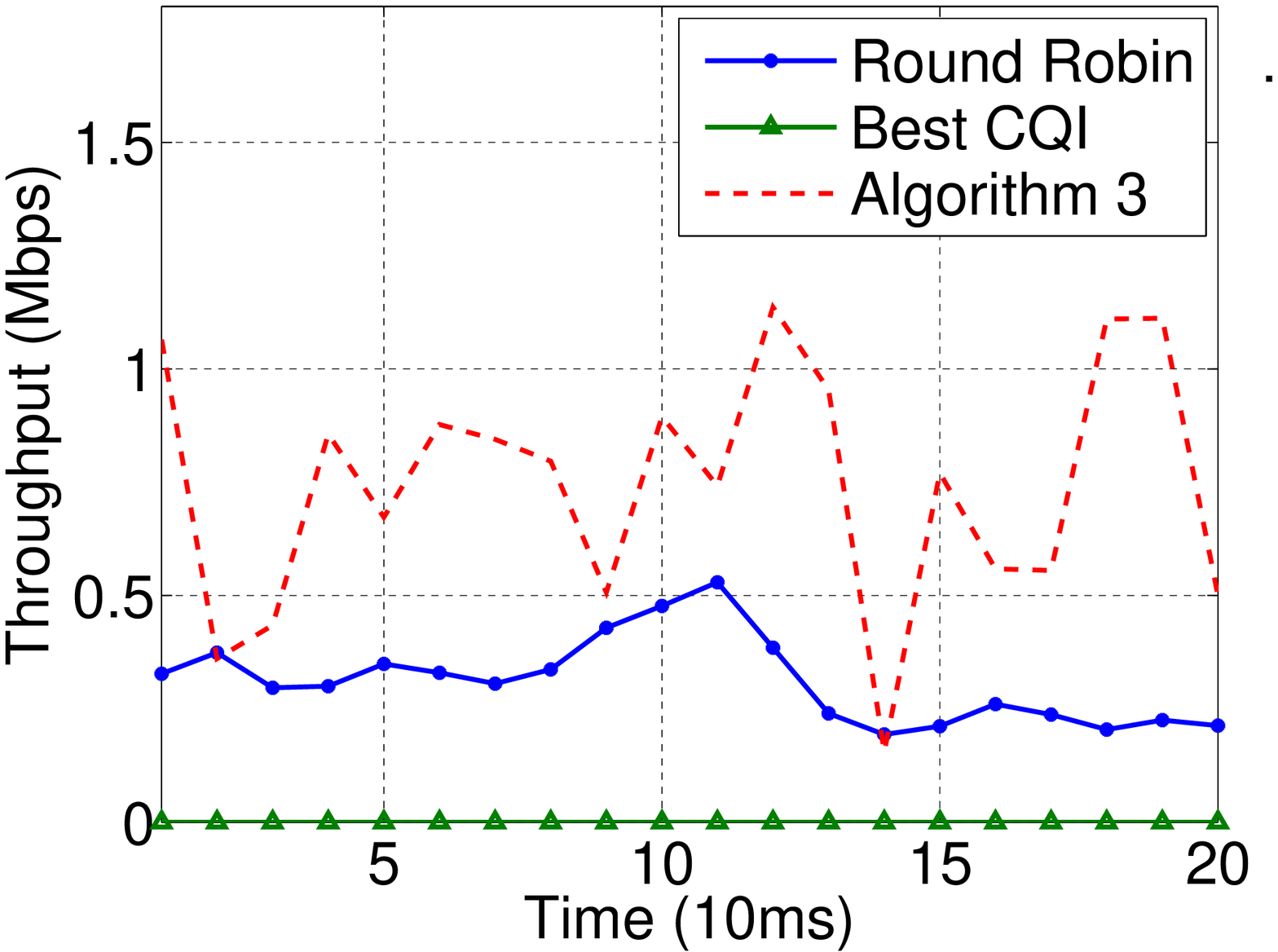}
    \caption{UE1's throughput comparison for $t = 1\sim 200 ms$.}
    \label{fig:rrbest-throughput2}
    \end{minipage}
    \hfill
    \begin{minipage}[t]{0.48\linewidth}
    \centering
    \includegraphics[width=\textwidth]{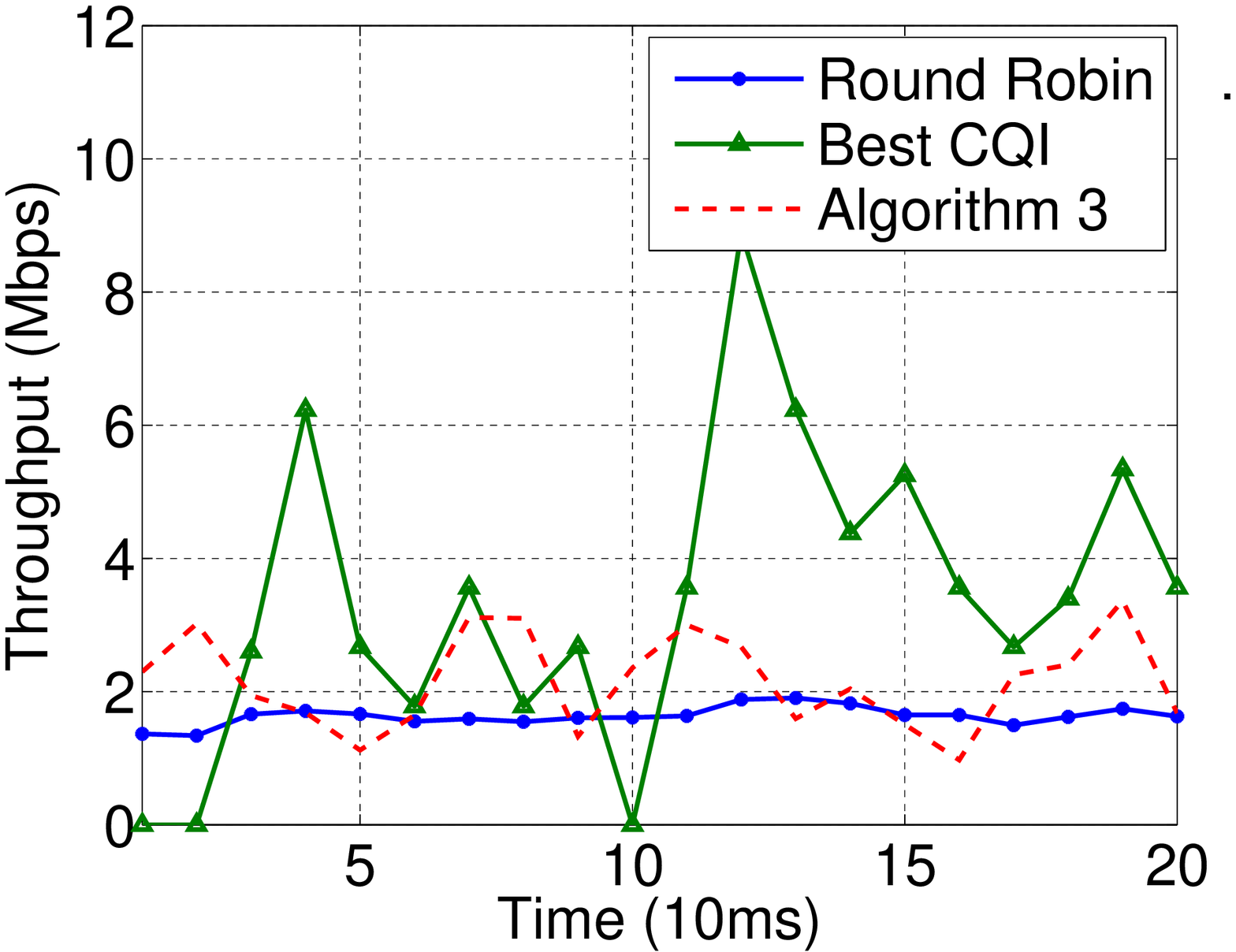}
    \caption{UE17's throughput comparison for $t = 1\sim 200 ms$.}
    \label{fig:rrbest-throughput3}
    \end{minipage}
   	\end {center}
\end{figure}

Figure \ref{fig:rrbest-throughput2} and Figure \ref{fig:rrbest-throughput3} further show that the high throughput of the Best CQI scheduler come at the expense of fairness. For example, UE1 has almost no throughput while UE17 has a rather high throughput for the first 200 $ms$ under the Best CQI scheduler. Most users experience low throughput under the best CQI scheduler. Our algorithm instead can distribute resource blocks relatively fairly. When a user needs high throughput to transmit its data, it could bid a higher price.

\section{Conclusion}
\label{sec:conclusion}

This paper discusses inband relaying solution for improving coverage without requiring extra spectrum resources for backhauling. We propose auction-based solutions, targeting at dynamic spectrum resources sharing and allocating the resources to serve users who value them most. Two truthful auctions are proposed for the relaying base station model and the extended HetNet model, repsectively. The first auction ensures an approximation ratio of $\frac{\delta-2}{\delta e -2}$ in terms of social welfare. The second auction focuses on tackling the heterogeneity of resource blocks. Extensive simulations are conducted for a larger scale scenario to verify the efficacy of proposed auctions, showing that auctions can improve both coverage and spectrum efficiency. 


\appendix
\label{sec:appendix}

\edef\refagainthmruntime{\ref{thm:alg1_runtime}}
\section{Proof of Theorem \refagainthmruntime}
\label{appendix:theorem3}

\noindent{\em Proof:} First we examine the runtime of Algorithm \ref{alg:greed}. The sorting operation can be performed in $O(|\mathcal{B}|\log (|\mathcal{B}|))$ using, {\em e.g.}, Merge-Sort. In each iteration, Algorithm \ref{alg:greed} picks a bid from $\mathcal{B}$. Therefore the {\tt while}-loop will run at most $|\mathcal{B}|$ times, which is polynomial to the input size. The loop body runs in constant time (line 7 also costs constant time since we can pick the first element and remove it when all bids are sorted in descending order at the very beginning). Therefore the upper bound of the runtime of Algorithm \ref{alg:greed} is $O(|\mathcal{B}|+|\mathcal{B}|\log (|\mathcal{B}|)) = O(|\mathcal{B}|\log (|\mathcal{B}|))$.

Since all resource blocks are considered identical, the primal constraints are transformed to the following:

{\small
\begin{equation}
\sum_{i=1}^{|\mathcal{B}|} r_ix_i + \max_{j\in\mathcal{B}_r} r_jx_j \leq |N|.
\end{equation}}

Let $\mu$ be the first bid violating the feasibility among the bids added to the solution set. Let $\mathcal{C}$ be the solution set before $\mu$ is added. We have 

{\small
\begin{eqnarray*}
&\sum_{i\in \mathcal{C}\cup\{\mu\}} r_i + \max_{j\in\mathcal{B}_r\cap (\mathcal{C}\cup\{\mu\})} r_j > |N|\\
\Leftrightarrow&\sum_{i\in \mathcal{C}} r_i + r_{\mu}+ \max_{j\in\mathcal{B}_r\cap (\mathcal{C}\cup\{\mu\})} r_j > |N|\\
\Rightarrow& \sum_{i\in \mathcal{C}} r_i > |N| - 2\max_{j\in\mathcal{B}} r_j\\
\Rightarrow& \sum_{i\in \mathcal{C}} r_i / (|N| - 2\max_{j\in\mathcal{B}} r_j) > 1.\\
\end{eqnarray*}}
which further implies that:
{\small
\begin{equation*}
\begin{split}
|N|\lambda &= \frac{1}{|N|}|N|\exp(\delta-2)^{\sum_{i\in\mathcal{C}} \frac{r_i}{|N| - 2\max_{j\in\mathcal{B}} r_j}} > \exp(\delta-2).\\
\end{split}
\end{equation*}}
\noindent which violates the termination condition, so $\mu$ will not be added to the solution set. Thus Algorithm \ref{alg:greed} provides a feasible solution to the primal problem. \qed

\edef\refagainthmappro{\ref{thm:alg1-appro}}
\section{Proof of Theorem \refagainthmappro}
\label{appendix:theorem4}

\noindent {\em Proof:} Let $d^*$ be the value of the optimal dual objective. Let $m = \max_{i\in\mathcal{B}} r_i$. Suppose $(v_{\mu}, r_{\mu})$ is the bid selected in iteration $t-1$. We also assume that the last iteration is $\omega$. According to the update rule of $\lambda$, we have:

{\small
\begin{equation*}
\begin{split}
|N|\lambda^t &= |N|\lambda^{t-1} \cdot \exp(\delta -2)^{r_{\mu}/(|N|-2m)}\\
&=|N|\lambda^{t-1}(1+(\exp(\delta -2)^{m/(|N|-2m)}-1))^{r_{\mu}/m}\\
&\leq |N|\lambda^{t-1}(1+ \frac{r_{\mu}}{m}(\exp(\delta -2)^{m/(|N|-2m)}-1))\\ 
&= |N|\lambda^{t-1} + \lambda^{t-1}\frac{r_{\mu}|N|}{m}(\exp(\delta -2)^{m/(|N|-2m)}-1),
\end{split}
\end{equation*}}
\noindent where the last inequality is due to $(1+a)^x\leq (1+ax), \forall x\in [0,1]$. 
Note that $\delta = |N|/\max_{i\in\mathcal{B}}r_i=|N|/m$, we have:

{\small
\begin{equation}
\label{eqn:dual-iteration}
\begin{split}
|N| \lambda^t &\leq |N| \lambda^{t-1} + \lambda^{t-1}r_{\mu} \delta (\exp(\delta-2)^{1/(\delta-2)}-1)\\
&= |N| \lambda^{t-1} + \lambda^{t-1}r_{\mu} \delta (e-1)\\
&= |N|\lambda^{t-1} + v_{\mu}\delta(e-1)/g(\lambda^{t-1}).
\end{split}
\end{equation}}
\noindent where $g(\lambda^{t-1}) = \frac{v_{\mu}}{\lambda^{t-1}r_{\mu}}$. 

Next the approximation ratio is analyzed as follows. It is trivial if Algorithm \ref{alg:greed} terminates due the condition $\mathcal{C} == \mathcal{B}$. Because all bids are accepted in that case, and the result is optimal. 

As a result, we assume that Algorithm \ref{alg:greed} terminates due to the condition $|N| \lambda^{\omega}\ > \exp(\delta -2)$. A lower bound on the solution obtained from Algorithm \ref{alg:greed} will be found.

We first show that a dual solution $(\lambda^{t-1}, \bm{\rho}^{t-1}, \bm{\xi}^{t-1})$ obtained from Algorithm \ref{alg:greed} can be converted into a feasible dual solution $\big(g(\lambda^{t-1})\lambda^{t-1}, g(\lambda^{t-1})\bm{\rho}^{t-1}, \bm{\xi}^{t-1}\big)$ by the following three cases.

{\em Case 1}, $\forall i\in \mathcal{C}\subseteq \mathcal{B}$, we know that 

{\small
\begin{equation*}
(r_i \lambda^{t-1}  + \rho_i^{t-1} \bm{1}_{i\in\mathcal{B}_r})g(\lambda^{t-1})+\xi_i^{t-1} - v_i \geq 0,
\end{equation*}}
holds for any $\lambda^{t-1}>0, \rho_i^{t-1}\geq 0$ since $\xi_i^{t-1} = v_i$.

{\em Case 2}, $\forall i\in \mathcal{B}\setminus\mathcal{C}$, $(r_i \lambda^{t-1} + \rho_i^{t-1} \bm{1}_{i\in\mathcal{B}_r})g(\lambda^{t-1})+\xi_i^{t-1} - v_i = r_i \frac{v_{\mu}}{r_{\mu}} + \rho_i^{t-1} \bm{1}_{i\in\mathcal{B}_r} \frac{v_{\mu}}{\lambda^{t-1}r_{\mu}} + \xi_i^{t-1} - v_i$. According to line 7 in Algorithm \ref{alg:greed}, we know that $\frac{v_{\mu}}{r_{\mu}} \geq \frac{v_i}{r_{i}}, \forall i\in \mathcal{B}\setminus\mathcal{C}$, therefore we have that: $r_i \frac{v_{\mu}}{r_{\mu}} \geq v_i$. Therefore we know that $(\ref{eqn:wdp-no-cqi-dual2}a)$ holds for the converted dual solution.

{\em Case 3}, Next we examine constraint $(\ref{eqn:wdp-no-cqi-dual2}b)$. 
{\small
\begin{equation*}
\begin{split}
\lambda^{t-1} g(\lambda^{t-1}) - \sum_{i\in\mathcal{B}_r} \rho_i g(\lambda^{t-1})
=& (\lambda^{t-1} - \sum_{i\in\mathcal{B}_r} \rho_i) g(\lambda^{t-1})\\
\geq& (\lambda^{t-1} - |\mathcal{B}_r| \frac{\lambda^{t-1}}{|\mathcal{B}_r|}) g(\lambda^{t-1})\geq 0,
\end{split}
\end{equation*}}
\noindent where the first inequality holds due to $\lambda^{t}$ increases as $t$ increases. Thus $(\lambda^{t-1} g(\lambda^{t-1}), g(\lambda^{t-1})\bm{\rho}^{t-1}, \bm{\xi}^{t-1})$ is a feasible solution to the dual problem in $(\ref{eqn:wdp-no-cqi-dual2})$. According to the weak duality, we have
{\small
\begin{equation*}
\begin{split}
d^* \leq |N| \lambda^{t-1} g(\lambda^{t-1}) + \sum_{i\in\mathcal{B}}\xi_i^{t-1}
\Longleftrightarrow  \frac{d^* - \sum_{i\in\mathcal{B}}\xi_i^{t-1}}{|N|\lambda^{t-1}} \leq g(\lambda^{t-1}).\\
\end{split}
\end{equation*}
}
Note that if $\exists t\leq \omega, \sum_{i\in\mathcal{B}} \xi_i^t/d^* \geq \alpha$, then we have $p \geq  \sum_{i\in\mathcal{B}} \xi_i^t \geq \alpha d^*$, which implies that the approximation ratio is $\alpha = \frac{\delta-2}{\delta e - 2}$. Now suppose $\forall t\leq \omega, \sum_{i\in\mathcal{B}} \xi_i^t/d^*  < \alpha$, then 
{\small
\begin{equation}
\label{eqn:fitting-func}
(1-\alpha) \frac{d^*}{|N|\lambda^{t-1}}\leq \frac{d^* - \sum_{i\in\mathcal{B}}\xi_i^{t-1}}{|N|\lambda^{t-1}} \leq g(\lambda^{t-1}).
\end{equation}
}
Combining (\ref{eqn:dual-iteration}) and (\ref{eqn:fitting-func}), we have:
{\small
\begin{equation*}
\begin{split}
|N|\lambda^{t} &\leq |N|\lambda^{t-1} + \frac{v_{\mu}\delta(e-1) |N|\lambda^{t-1}}{(1- \alpha)d^*}\\
& = |N|\lambda^{t-1}(1+\frac{v_{\mu}\delta(e-1)}{(1-\alpha)d^*})\\
&\leq |N|\lambda^{t-1}\exp(\frac{v_{\mu}\delta(e-1)}{(1-\alpha)d^*}).\\
\end{split}
\end{equation*}
}
Apply the above inequality for $t=0,...,\omega$:
{\small
\begin{equation*}
|N|\lambda^{\omega} \leq |N|\frac{1}{|N|} \exp(\frac{p\delta(e-1)}{(1-\alpha)d^*}).
\end{equation*}
}
Next recall that Algorithm \ref{alg:greed} terminates due to the condition $|N| \lambda^{\omega}\ > \exp(\delta -2)$, then we have
$\exp(\delta-2) < \exp(\frac{p\delta(e-1)}{(1-\alpha)d^*})$,  which implies 
$\alpha = \frac{(\delta - 2)(1-\alpha)}{\delta(e-1)} \Longleftrightarrow \alpha  = \frac{\delta-2}{\delta e-2}$. \qed

\end{document}